\documentclass[11pt,a4paper]{article}
\pdfoutput=1
\usepackage{jheppub}
\usepackage{gensymb}
 \usepackage{multirow}
 \usepackage[table,xcdraw]{xcolor}
\DeclareSymbolFont{matha}{OML}{txmi}{m}{it}
\DeclareMathSymbol{\varv}{\mathord}{matha}{118}
\usepackage{amsmath}
\usepackage{amssymb}
\usepackage{graphicx}
\usepackage{subfigure}
\usepackage{pstricks}
\usepackage{bm}
\usepackage{pbox}
\usepackage{placeins}
\usepackage{booktabs}
\usepackage[T1]{fontenc}
\usepackage{footnote}
\usepackage{pdfpages}
\usepackage{hhline}
\usepackage{multirow}
\usepackage{multicol}
\usepackage{enumitem}
\usepackage[toc,page]{appendix}
\allowdisplaybreaks
\usepackage{comment}
\usepackage{float}                                             
\usepackage{mathtools}                  
\usepackage{textcomp}
\usepackage{gensymb}
\usepackage{relsize}
\usepackage[numbers]{natbib}
\usepackage{notoccite}
\usepackage{rotating}
\usepackage{tabularx}
\usepackage[labelfont=bf]{caption} 
\FloatBarrier
\usepackage[many]{tcolorbox}

\renewcommand{\thetable}{\Roman{table}}

\definecolor{MyDarkBlue}{rgb}{0.1, 0.1, 0.8} 
\definecolor{SBlue}{rgb}{0.2, 0.4, 0.7} 
\definecolor{MyLightBlue}{rgb}{0.22,0.51,0.9}
\definecolor{MyGreen}{rgb}{0.0, 0.5, 0.0}
\definecolor{BrickRed}{rgb}{0.8, 0.25, 0.33}
\RequirePackage{hyperref}
\hypersetup{colorlinks, citecolor=MyGreen,linkcolor=MyDarkBlue, urlcolor=BrickRed}

\begin{document}

\title{\LARGE Vector Boson Dark Matter From Trinification
 }
 \author[a]{K.S. Babu,}
 \author[b]{Sudip Jana,}
 \author[c]{Anil Thapa}
 
 \affiliation[a]{Department of Physics, Oklahoma State University, Stillwater, OK 74078, USA}
 \affiliation[b]{Max-Planck-Institut für Kernphysik, Saupfercheckweg 1, 69117 Heidelberg, Germany}
 \affiliation[c]{Department of Physics, University of Virginia, Charlottesville, Virginia 22904-4714, USA}

\emailAdd{babu@okstate.edu, sudip.jana@mpi-hd.mpg.de,  wtd8kz@virginia.edu}
\abstract{We show how
trinification models based on the gauge group $SU(3)_C \times SU(3)_L \times SU(3)_R$ realized near the TeV scale can provide naturally a variety of dark matter (DM) candidates. These models contain a discrete $T$ parity which may remain unbroken even after spontaneous symmetry breaking.  The lightest $T$-odd particle, which could be a fermion, a scalar, or a gauge boson, can constitute the dark matter of the universe. This framework naturally admits a doublet-singlet fermionic DM, a  singlet scalar DM, or a vector boson DM. Here we develop the vector boson DM scenario wherein the DM couples off-diagonally with the usual fermions and vector-like fermions present in the theory. We show consistency of this framework with dark matter relic abundance and direct detection limits as well as LHC constraints. We derive upper limits of 900 GeV on the vector gauge boson DM mass and 4.5 TeV on the vector-like quark masses. We also show the consistency of spontaneous gauge symmetry breaking down to Standard Model times an extra $U(1)$ while preserving the $T$-parity. 

}


\maketitle
\section{Introduction}\label{sec:Intro}
Models based on the gauge symmetry $SU(3)_C \times SU(3)_L \times SU(3)_R$, termed trinification \cite{Rujula:1984}, were proposed as grand unified theories where matter fields as well as the forces acting on them are unified \cite{Rujula:1984,Babu:1985gi,Sayre:2006ma,Hetzel:2015bla,Kim:2004pe}.  This is possible owing to the three identical $SU(3)$ factor groups by imposing a cyclic permutation symmetry among them.  The three gauge couplings will then all be equal at the scale where the trinification symmetry is realized. This scale turns out to be of order $10^{14}$ GeV in such a framework as can be inferred by extrapolating the three gauge couplings of the Standard Model (SM) to higher energies. 

The trinification gauge structure may be interesting in itself, without necessarily imposing the cyclic permutation symmetry.  In this scenario trinification can be realized around the TeV scale, which would make it accessible directly to collider experiments \cite{Pelaggi:2015kna,Babu:2017xlu,Wang:2018yer}.  Many of the merits of the unified framework would be preserved in this setup, including the quantization of electric charge (sine the gauge symmetry is non-Abelian), structure of matter multiplets and inter-relations among fermion masses. These models are asymptotically free and can be extrapolated all the way up to the Planck scale \cite{Pelaggi:2015kna,Wang:2018yer}, in contrast to  the case in the SM.  Unlike unified theories based on simple groups such as $SU(5)$, the trinification symmetry group does not have gauge boson mediated proton decay, which facilitates its realization near the TeV scale. It is this class of models with TeV scale realization that is the focus of this paper.

The main goal of this paper is to show how a variety of dark matter candidates arises naturally within the trinification framework, and develop the phenomenology of one such scenario wherein the DM candidate is a vector gauge boson.  Natural identification of DM candidates is possible owing to a discrete trinification parity ($T$-parity) that remains unbroken even after spontaneous symmetry breaking.  The $T$ parity is defined as
\begin{equation}
    T = (-1)^{I_{8_L} + I_{8_R} + 2 S}  \, ,
    \label{eq:Tparity}
\end{equation}
where $I_{8_L}$ and $I_{8_R}$ are the charges (generalized isospins) under the diagonal generators $T_{8L}$ of $SU(3)_L$ and $T_{8R}$ of $SU(3)_R$ and $S$ is the spin of the particle.  Previous studies of symmetry breaking in trinification models  broke this $T$ parity  \cite{Babu:1985gi,Sayre:2006ma,Hetzel:2015bla,Kim:2004pe,Pelaggi:2015kna,Babu:2017xlu}. Here we show by explicit construction how to consistently achieve  symmetry breaking while preserving this $T$-parity.

The types of DM candidates that can arise within this framework are few and thus predictive. The anomaly-free fermion representations of the theory contain new quarks and leptons which transform chirally under the full symmetry group, but vectorially under the SM gauge symmetry. These new vector-like fermions are $T$-odd, and the lightest neutral particle among them may be identified as the DM candidate. The quantum numbers of these fermions are fixed by the gauge structure, and thus the nature of such a fermionic DM candidate is fully determined.  It turns out that this candidate is a mixed doublet-singlet fermion under the weak $SU(2)_L$ symmetry.  Unlike the usually discussed doublet-singlet fermion DM models \cite{Cohen:2011ec,Calibbi:2015nha,Abe:2019wku,Barman:2019aku}, here the DM mass is protected by a gauge symmetry. Such a  chiral fermion DM is more appealing \cite{Dev:2016qeb,Berryman:2017twh,Abe:2019zhx}.  Since this scenario has been well studied in the literature, here we do not discuss it any detail, except to note that there would be some differences in its phenomenology owing to couplings of the DM fermion with additional particles (gauge bosons and Higgs bosons) that are present in the theory. In the next section we  identify this fermionic DM candidate and briefly comment on the correlation of its mass with the mass of the right-handed neutrino involved in the seesaw mechanism.

The symmetry breaking sector, which we develop in detail here, also has $T$-odd scalars, the lightest of which may be a dark matter candidate.  These particles are either singlets, doublets, or triplets of the weak $SU(2)_L$ gauge group. This restriction arises since we use only two types of Higgs fields for symmetry breaking with quantum numbers $(1,3,3^*)$ and $(1,6,6^*)$.  The first field is needed to generate quark and lepton masses, including the vector-like fermion masses, and the second is needed to realize the seesaw mechanism. 
Thus the DM candidate could be a linear combination of three multiplts -- singlet, doublet and triplet -- of weak $SU(2)_L$.  However, as we show by an explicit Higgs potential analysis, among these scalars, only a $T$-odd singlet survives down to the TeV scale, if the trinification symmetry breaks at a higher scale down to the SM times a $U(1)$.  There are also two $T$-even singlet scalars at or below the TeV scale, into which this $T$-odd DM scalar can annihilate.  Thus this scenario is very similar to complex singlet DM models where a stable scalar singlet annihilates into its unstable scalar partners \cite{Gonderinger:2012rd,Coito:2021fgo}. If the trinification symmetry breaks down directly to the SM, a combination of scalar singlet, doublet and triplet may be made light to serve the role of DM.

Fermionic or bosonic dark matter candidates which are naturally stable have been studied in the context of the SM with an extended particle spectrum \cite{Cirelli:2005uq}, left-right symmetric theories \cite{Heeck:2015qra},  and unified theories based on $SO(10)$ \cite{Kadastik:2009cu,Mambrini:2015vna}.  In these models, typically new fields are introduced to serve the role of DM, with their stability ensured by remnants of the  gauge symmetry.  In the trinification framework, these new particles are already present within the minimal setup, along with the $T$ parity to stabilize them.  This framework can be embedded into $E_6$ unified theories while preserving the $T$ parity \cite{Babu:2021}. 

Our main focus here is however a $T$-odd gauge boson dark matter candidate that is available in the theory.  It is electrically neutral and also a singlet of the SM gauge symmetry.  It has the feature that it couples off-diagonally to fermions and scalars.  For example, it connects the $T$-even SM fermions with the $T$-odd vector-like fermions, and similarly in the scalar sector. While the trinification symmetry group has several new gauge bosons, the DM candidate we identify originates from the symmetry breaking chain
\begin{eqnarray}
    SU(3)_C \times SU(3)_L \times SU(3)_R &\rightarrow& SU(3)_C \times SU(2)_L \times U(1)_Y \times O(2)_R \nonumber \\
    &\rightarrow& SU(3)_C \times SU(2)_L \times U(1)_Y~.
    \label{sym}
\end{eqnarray}
Here at the first stage of symmetry breaking an $O(2)$ symmetry is left unbroken, along with the SM gauge symmetry.  The gauge boson of this $O(2)$ is the dark matter candidate that we analyze in detail. Although the $O(2)$ symmetry is isomorphic with a $U(1)$, we denote it as $O(2)$ in order to emphasize the off-diagonal nature of its couplings to fermions. The first stage of symmetry breaking in Eq. (\ref{sym}) is achieved by a Higgs field transforming as $(1,6,6^*)$ under the trinification gauge group.  This Higgs multiplet also generates Majorana masses for the right-handed neutrinos via renormalizable couplings.  The second stage of symmetry breaking is mediated via Higgs fields which transform as $(1,3,3^*)$ under trinification. These fields are needed to also break the electroweak symmetry and to provide masses to quarks and leptons.
Our analysis shows that this vector boson DM candidate is consistent with relic abundance constraints, direct detection limits, as well as LHC constraints.  In fact, we find an upper limit of 900 GeV on its mass, along with upper limits of 4.5 TeV and 6 TeV on the vector-like quark and lepton masses of the model.  The upper limits on vector-like fermion masses arise from perturbative unitarity -- their masses arise from the same Higgs field that generates the vector boson DM candidate a mass. The entire parameter space of the model would be probed by future colliders.  

Vector boson DM candidates have been studied in several other contexts \cite{Diaz-Cruz:2010czr,Chen:2009ab,Abe:2020mph,Barman:2017yzr,Maru:2018ocf,Chowdhury:2021tnm}.  The trinification DM differs from most models in this class in that 
the gauge interactions of the DM in our framework is completely determined in terms of $SU(2)_L \times U(1)_Y$ gauge couplings. This makes  the model more constrained and predictive. The spectrum of particles at or around the TeV scale in the model we analyze consists of the vector boson DM, three families of vector-like quarks and leptons, a new Higgs boson $h'$ which is a remnant of the $O(2)_R$ symmetry breaking, a $T$-odd scalar singlet $H'$, and a $T$-even scalar singlet $\tilde{h}$ as well as a doubly charged $T$-even  scalar $\delta^{++}$.  The last three particles in this list arise as a result of consistency with symmetry breaking within the minimal setup.

The rest of the paper is organized as follows.  In Sec. \ref{sec:model} we describe the trinification model and its symmetry breaking pattern. The gauge boson masses and the Higgs potential are constructed here, with details of the Higggs spectrum delegated to an Appendix.  In Sec. \ref{sec:LHC} we analyze LHC constraints on vector-like leptons and quarks, with their signature decays into leptons/quarks plus missing transverse energy.  In Sec. \ref{sec:DMrelic} we analyze the dark matter relic abundance and identify parameter space  consistent with  direct detection and LHC constraints. In Sec. \ref{sec:con} we conclude.

\section{TeV Scale Trinification Model}\label{sec:model}
The trinification model is based on the gauge symmetry $SU(3)_C \times SU(3)_L \times SU(3)_R$ under which quarks and leptons fields are assigned as
\begin{equation}
    Q_{L}\left(3,3^{*}, 1\right)=\left(\begin{array}{l}
u \\
d \\
D
\end{array}\right)_{L}, \quad Q_{R}\left(3,1,3^{*}\right)\left(\begin{array}{l}
u \\
d \\
D
\end{array}\right)_{R}, \quad \psi_{L}\left(1,3,3^{*}\right)=\left(\begin{array}{ccc}
E^{0} & E^{-} & e^{-} \\
E^{+} & E^{c 0} & \nu \\
e^{c} & \nu^{c} & N
\end{array}\right)_L \, .
\label{eq:fermions}
\end{equation}
Here $D_L$ and $D_R$ are new vector-like iso-singlet quarks under the SM, while ($E^0, E^+)$ and $(E^-, E^{c0}$) are vector-like lepton doublets with hypercharge $Y = +1/2$ and $-1/2$ respectively. $\nu^c$ is the conjugate of the right-handed neutrino, while $N$ is a new neutral singlet of the SM. The $SU(2)_L$ and $SU(2)_R$ subgroups in Eq.~\eqref{eq:fermions} respectively run vertically and horizontally. Note that $e^c$ and $e^-$ are grouped in the same gauge multiplet which implies that lepton number is explicitly broken in the model.

To break the trinification symmetry spontaneously all the way down to $SU(3)_C \times U(1)_{em}$, we employ three copies of bi-triplet  Higgs fields $\Phi_n \sim (1,3,3^*)$  and a bi-sextet scalar field $\chi \sim (1, 6, 6^*)$. Three copies of the $\Phi$ fields are used so that there are no obvious relations among the fermion masses, which could  potentially go wrong.  The field content of the $\Phi_n$  is analogous to the lepton field in Eq.~\eqref{eq:fermions}. We denote its components as $\Phi_i^\alpha$. The bi-sextet field $\chi$ is represented tensorially as $\chi_{ij}^{\alpha\beta}$ and its components can be identified from the coupling $\Phi \Phi \chi^*$  which is gauge invariant. The vacuum expectation values (VEVs) of these Higgs fields are denoted as
\begin{equation}
    \left\langle\Phi_{n}\right\rangle=
           \left(\begin{array}{ccc}
              v_{u n} & 0 & 0 \\
              0 & v_{d n} & 0 \\
              0 & 0 & V_{n}
          \end{array}\right) \, , \quad \quad 
          \langle \chi^{22}_{33} \rangle = V_\nu \, , \quad \quad
          \langle \chi_{33}^{33} \rangle = V_N \, .
          \label{eq:vevs}
\end{equation}
We shall assume for simplicity in our analysis that these VEVs are real.  
Note that the $\langle \Phi_2^3\rangle$, the VEV of the $\nu^c$-like scalar, see Eq. (\ref{eq:fermions}), could be nonzero in general, but if it is zero, an unbroken $T$ parity can be realized.  In models with multiple $\Phi_n$ fields and no analog of the $\chi$ field, trinification symmetry breaking down to the SM would require some $\langle \Phi_2^3\rangle$ to be nonzero, which would break the $T$-parity. 
With only the $\Phi_n$ fields acquiring VEVs as in Eq. (\ref{eq:vevs}), trinification symmetry breaks down to $SU(3)_C \times SU(2)_L \times SU(2)_R \times U(1)$, suggesting the need for an additional scalar field. The bi-sextet field $\chi$ serves this role. From the $\chi$ field, $\langle\chi_{23}^{33}\rangle$, the VEV of the $(\nu^c N)$ component, could be nonzero in general, but $T$ parity can be preserved if it is zero.  Only the $(\nu^c\nu^c)$ and $(NN)$ components of $\chi$ acquire VEVs, but this is sufficient for reducing the gauge symmetry down to the SM. This is the main difference in our construction compared to other symmetry breaking patterns in trinification \cite{Rujula:1984,Babu:1985gi,Sayre:2006ma,Hetzel:2015bla,Kim:2004pe,Pelaggi:2015kna}. 

We note that the $T$ parity of a general tensor representation $ X_{ijk ...}^{\alpha \beta \gamma ....}$
is given by $(-1)^{I_i+I_j+I_k+....+I_\alpha+ I_\beta + I_\gamma + .... + 2S}$ where $I_i = 1$ for $i=1,2$ and $I_i = -2$ for $i=3$ are the $T_8$ eigenvalues. Under $T$-parity, all SM fermions are then even, including the $\nu^c$ field, while the vector-like fermions $D_L$, $D_R$, ($E^0, E^+)$ and $(E^-, E^{c0}$) are odd. A similar classification can be made for the scalar fields as well. 

The VEVs of $\chi$, $V_\nu$ and $V_N$, can be chosen to be equal and much larger than the VEVs of $\Phi_n$. In this case we can identify the two stage symmetry breaking shown in Eq. (\ref{sym}). The identification $V_N = V_\nu$ is supported by the Higgs potential, as we show below. Once the smaller $\Phi_n$ VEVs develop, the equality $V_N =V_\nu$ can no longer be preserved, owing to an induced value of $V_N - V_\nu$ which could be of order $V_n$, the smaller VEV. This is the scenario that we investigate which yields a TeV scale vector boson dark matter candidate, with all other gauge bosons being heavier than a few TeV.

\subsection{Fermion mass generation}
There are 27 Weyl fermions per generation of \{$Q_L \oplus Q_R \oplus \psi_L$\} of which 15 are the usual SM chiral fermions, the $\nu^c$ field is the singlet fermion involved in the seesaw mechanism, and the remaining 10 fields are vector-like iso-singlet quarks and iso-doublet leptons. As noted earlier, in order to generate the observed pattern of quark and lepton masses, three copies of $\Phi_n \sim (1,3,3^*)$ Higgs fields and one $\chi \sim (1, 6, 6^*)$ field are introduced. The most general Yukawa interactions of quark and leptons with these Higgs fields are given by
\begin{equation}
    -{\cal L}_Y = Y_{qn} \Bar{Q}_{L \alpha} (\Phi_{n})_i^\alpha Q_{R}^i + Y_{\ell n}\ \psi_i^\alpha \psi_j^\beta (\Phi_{n})_k^\gamma \epsilon^{ijk} \epsilon_{\alpha \beta \gamma} + y_{\ell}\ \psi_i^\alpha \psi_j^\beta \chi^{ij}_{\alpha \beta} + h.c. \, ,
    \label{eq:LagY}
\end{equation}
where ($i,j,k$) are $SU(3)_L$ indices,  ($\alpha, \beta,\gamma$) are $SU(3)_R$ indices, and $Y_{qn}, Y_{\ell n}$ and $y_\ell$ are $3\times 3$ Yukawa coupling matrices, with $y_\ell$ being symmetric in family indices, which we have suppressed here. Here a summation over $n=1, 2, 3$ is assumed. Expanding Eq. (\ref{eq:LagY}) in component form the mass matrices for up-type, down-type and vector-like quarks are found to be
\begin{equation}
    M_u = Y_{qn}\ v_{un} \, , \quad M_d = Y_{qn}\ v_{dn} \, , \quad M_D = Y_{qn}\ V_n \, .
    \label{eq:exoticquark}
\end{equation}
The need for multiple copies of $\Phi_n$ fields is clear from here. With a single $\Phi$ field all three matrices would become proportional, resulting in no quark mixing.  Even with two copies of $\Phi$, the vector-like quark mass matrix can be derived from those of the light quarks, which does not support TeV scale realization of trinification \cite{Babu:2017xlu}. The vector-like quark masses would be hierarchical in this case, suggesting that trinification symmetry is broken at a scale much larger than a few TeV. With three copies of $\Phi_n$, all fermion masses and mixings can be accommodated, along with TeV scale symmetry breaking. One could, for example, realize a scenario where the up-type quark masses arise from $\Phi_1$, the down-type masses from $\Phi_2$, and the vector-like quark masses from $\Phi_3$, which would allow for hierarchical Yukawa couplings in the up- and down-type mass matrices, along with non-hierarchical couplings in the vector-like quark mass matrix. 

The mass matrices for the charged leptons are given by
\begin{equation}
    M_e = -Y_{\ell n}\ v_{dn} \, , \quad M_{E^\pm} = -Y_{\ell n}\ V_n \, .
    \label{eq:chargeL}
\end{equation}
Having three copies of $\Phi_n$ would give the flexibility to  give independent masses to the leptons and the vector-like leptons as well. In the neutral lepton sector the mass matrices decouple into a $2 N_g\times 2N_g$ $T$-even matrix and a $3 N_g \times 3 N_g$ $T$-odd matrix where $N_g$  is the number of generations, which are given by
\begin{equation}
    M_\nu = \begin{pmatrix}
            0 &~~~ - Y_{\ell n} v_{un}\\
            - Y_{\ell n} v_{un} &~~~ y_{\ell} V_\nu 
            \end{pmatrix} \, , \hspace{10 mm}
        M_N = \begin{pmatrix}
              0  &~~ Y_{\ell n} V_n &~~ Y_{\ell n} v_{un} \\
              Y_{\ell n} V_n &~~ 0 &~~ Y_{\ell n} v_{dn} \\
              Y_{\ell n} v_{un} &~~ Y_{\ell n} v_{dn} &~~ y_{\ell} V_N
              \end{pmatrix} \, . 
              \label{eq:neutaralM}
\end{equation}
Here $M_\nu$ spans the basis $(\nu, \nu^c)$ and $M_N$ spans the basis $(E^{c0}, E^0, N)$. Note that in the limit of electroweak symmetry, obtained by setting $v_{un}$ and $v_{dn}$ to zero in Eq. (\ref{eq:neutaralM}), the field $N$ decouples from the rest, while the masses of $E^0$ and $E^+$ become degenerate, as they form an $SU(2)_L$ doublet.  The light neutrino mass matrix can be obtained from  the matrix $M_\nu$ of Eq.~\eqref{eq:neutaralM} by the seesaw formula with $Y_{\ell n} v_{un} \ll y_\ell V_\nu$ as
\begin{equation}
    m_\nu^{\rm light} \simeq -(Y_{\ell n} v_{un}) (y_\ell V_\nu)^{-1} (Y_{\ell n} v_{un}) \, .
\end{equation}
The mass matrix for the heavier state is simply $y_\ell V_\nu$. We shall set $N_g=3$ in our analysis.  

The decoupling of the two matrices in Eq. (\ref{eq:neutaralM}) is due to $T$ parity under which $\nu$ and $\nu^c$ are even, while $(E^{c0}, E^0, N)$ are odd. The lightest of these $T$-odd states can be a dark matter candidate. This is the doublet-singlet DM scenario studied extensively in the literature \cite{Cohen:2011ec,Calibbi:2015nha,Abe:2019wku,Barman:2019aku}. As already noted, one difference here is the chiral nature of the DM fermion under the full trinification symmetry which protects its mass.  Stability of the DM is guaranteed by a discrete gauge symmetry, the $T$ parity of the model.  Furthermore, the doublet-singlet DM mass matrix is closely related to the neutrino mass matrix as seen from Eq. (\ref{eq:neutaralM}).  Note in particular that the Majorana mass matrices of the $\nu^c$ and $N$ fields are proportional. Since we are interested in the case where $V_N \simeq V_\nu \gg V_n$, from the $T$-odd neutral lepton mass matrix it is possible to integrate out the $N$ field, which would provide a small Majorana mass correction to the nearly degenerate Dirac fermion doublet.  Since this Majorana mass should be of at least of order 100 keV to be consistent with DM direct detection limits, we conclude that $V_N$ cannot be many orders of magnitude larger than $V_n$ in this realization.

\subsection{Hierarchy of vacuum expectation values}

The three VEVs given on Eq. (\ref{eq:vevs}), which are singlets of the SM, can all be of the same order, or they can exhibit various hierarchies.  If the three are all of the same order, trinification symmetry breaks down directly to the SM.  Such a scenario would be consistent with a fermionic doublet-singlet DM candidate.  Even if these VEVs are of order 10 TeV, the correct relic abundance of DM can be realized by choosing some of the Yukawa couplings $y_\ell$ to be of order 0.1 in Eq. (\ref{eq:LagY}), so that the DM mass would be of order 1 TeV. One could also realize scalar DM candidate in this scenario, where a combination of $SU(2)_L$ singlet, doublet and triplet scalars is taken to have mass of order few hundred GeV.
This scenario with non-hierarchical VEVs would however not support vector boson DM candidate, since LHC limits on $W_R$ and $Z_R$ masses would push the vector boson DM mass to above 4 TeV, which would be incompatible with its relic abundance. 

The VEV $V_n$ of Eq. (\ref{eq:vevs}) breaks $SU(3)_L \times SU(3)_R$ down to $SU(2)_L \times SU(2)_R \times U(1)$, as does the VEV $V_N$.  On the other hand, the VEV $V_\nu$ breaks the symmetry down to $SU(2)_L \times SU(2)' \times U(1)$ where the $SU(2)'$ is a different subgroup.  Jointly these VEVs break the symmetry down to $SU(2)_L \times U(1)_Y$. In addition, when $V_N = V_\nu$, the surviving symmetry is $SU(2)_L \times U(1)_Y \times O(2)_R$. This allows for the following possible hierarchies in these VEVs: (i) $V_N \gg V_n, V_\nu$, (ii) $V_n \gg V_N, V_\nu$, (iii) $V_\nu \gg V_N, V_n$, and (iv) $V_N = V_\nu \gg V_n$.  The first three cases are compatible with a fermionic doublet-singlet DM candidate, or a scalar singlet-doublet-triplet DM candidate, but not a vector boson DM candidate for reasons given in the previous paragraph.  It is case (iv) that generates naturally a vector boson DM.  

The Higgs potential analysis that we carry out is very general, and all these hierarchies can be implemented in our results.  As we focus on the vector boson DM candidate here, the hierarchy (iv) will be our main focus. Note that none of the three SM singlet VEVs can be set to zero.  If $V_n = 0$, the vector-like fermion masses would become zero, and if $V_N = 0$, the neutral fermion $N$ would remain massless.  And $V_\nu \neq 0$ is needed for completing the gauge symmetry breaking down to $SU(2)_L \times U(1)_Y$. Although the extremum of the potential allows for these zero VEV solutions, see Eq. (\ref{eq:C3}), we shall discard them for consistency of the framework.
\subsection{Gauge boson sector}\label{sec:gauge}

In this section, we examine the gauge boson sector and obtain all the mass eigenvalues and eigenstates explicitly. In particular, we find that one of the 12 new gauge bosons can be lighter than the rest and be a potential DM candidate. 

 At the scale of $SU(3)_L \times SU(3)_R$ unification the gauge couplings obey the boundary condition
\begin{equation}
    \alpha_{R}^{-1} = \frac{3}{4} \alpha_{Y}^{-1} - \frac{1}{4} \alpha_{L}^{-1}~.
\end{equation}
Here $\alpha_L= g_L^2/(4\pi)$ and $\alpha_R = g_R^2/(4\pi)$ are the $SU(3)_L$ and $SU(3)_R$ fine-structure constants, while $\alpha_Y = g_Y^2/(4\pi)$ is the fine-structure constant corresponding to hypercharge. The value of $\alpha_L$ is identical to that of $g^2/(4\pi)$ evaluated at the scale of trinification embedding. The couplings $\alpha_L$ and $\alpha_R$ at a given energy scale can be calculated from the measured values of the low energy gauge couplings at $\mu = m_Z$. We take as input $\alpha_{em} (m_Z) = 1/127.940$ and $\sin^2 \theta_W (m_Z) = 0.23126$, which give $\alpha_L (m_Z) = 0.033798$ and $\alpha_Y (m_Z) = 0.010675$. Here $\alpha_{em}^{-1} = \alpha_{L}^{-1} + \alpha_{Y}^{-1}$ is used to obtain the value of $\alpha_Y$. We then use one-loop renormalization group equations to evaluate $\alpha_i$ to higher energy scales, with the beta function coefficients given by
$b_i = 41/6~ (-19/6)$ for $\alpha_{Y(L)}$ with contributions arising only from the SM particles. With these, we find that the ratio of $\alpha_R/\alpha_L = 0.49$ with $\alpha_L\ (5 \,{\rm TeV}) = 0.0355$ and $\alpha_Y\ (5\, {\rm TeV})= 0.0103$, leading to $\alpha_R(5\,{\rm TeV}) = 0.0174$. In our numerical studies we use these as input values.  

The fields $\Phi_n$ transform as $\Phi_n \to U_L \Phi_n U_R^\dagger$ under gauge symmetry.  The transformation of the $\chi$ field can be obtained from here in tensorial form. The covariant derivatives for the Higgs fields can then be written as
\begin{equation}
    {\cal L}_{gauge} = \sum_n D^\mu (\Phi_n)_i^\alpha\  D_\mu (\Phi_n)_\alpha^i  + \sum_{\alpha \beta i j} (D_\mu \chi_{ij}^{\alpha \beta})\ (D^\mu \chi^{ij}_{\alpha \beta}) \, ,
\end{equation}
where 
\begin{align}
    D^{\mu} (\Phi_{n})_i^\alpha &= \partial^{\mu} (\Phi_{n})_i^\alpha-\frac{i g_{L}}{2}\ (\vec{T} \cdot \vec{W}_{L}^{\mu})_i^k (\Phi_{n})_k^\alpha +\frac{i g_{R}}{2}\  (\vec{T} \cdot \vec{W}_{R}^{\mu})_k^\alpha\ (\Phi_{n})^k_i \, , \\
    D^\mu \chi_{ij}^{\alpha \beta} &= \partial^\mu \chi_{ij}^{\alpha \beta} - \frac{i g_L}{2} (\vec{T} \cdot \vec{W}_{L}^{\mu})^{k}_i\ \chi_{kj}^{\alpha \beta} - \frac{i g_L}{2} (\vec{T} \cdot \vec{W}_{L}^{\mu})^{k}_j\ \chi_{ik}^{\alpha \beta} \nonumber \\
    & ~~~~~~~~~~~~ + \frac{i g_R}{2} (\vec{T} \cdot \vec{W}_{R}^{\mu})_\gamma^\alpha\ \chi_{ij}^{\gamma \beta} + \frac{i g_R}{2} (\vec{T} \cdot \vec{W}_{R}^{\mu})_\gamma^\beta\ \chi_{ij}^{\alpha \gamma} \, .
\end{align}
We define the gauge boson multiplets in the $(1,8,1)_L$ and $(1,1,8)_R$ representation as
\begin{equation}
    \vec{T} \cdot \vec{W}_{L, R}^{\mu}=\left(\begin{array}{ccc}
W_{3}+\frac{W_{8}}{\sqrt{3}} & \sqrt{2} W^{+} & \sqrt{2} V^{+} \\
\sqrt{2} W^{-} & -W_{3}+\frac{W_{8}}{\sqrt{3}} & \sqrt{2} V^{0} \\
\sqrt{2} V^{-} & \sqrt{2} V^{0 *} & \frac{-2 W_{8}}{\sqrt{3}}
\end{array}\right)_{L, R}^{\mu} \, 
\end{equation}
with the following definitions:
\begin{equation}
    W^{\pm \mu} = \frac{W_1^\mu \mp i W_2^\mu}{\sqrt{2}}, \hspace{5mm} V^{\pm \mu} = \frac{W_4^\mu \mp i W_5^\mu}{\sqrt{2}}, \hspace{5mm}  V^{0(*)\mu} = \frac{W_6^\mu \mp i W_7^\mu}{\sqrt{2}}~. 
    \label{eq:gaugeV}
\end{equation}
Here $(V^{\mu+}, V^{\mu0})_{L,R}$ form $SU(2)_{L,R}$ doublets, while $(W_8^\mu)_{L,R}$ are singlets. Note that with the choice of our VEV pattern given in Eq. (\ref{eq:vevs}) which leads to $T$ parity conservation, there is no mixing between $W^{\mu \pm}$ and $V^{\mu \pm}$ fields which are even and odd respectively under $T$. 

The $T$-even charged gauge boson mass matrix reads, in the basis $(W_L^{\mu +} , W_R^{\mu +})$, as
\begin{equation}
    M_{W_{L,R}^+}^{2} = \frac{1}{2}
    \begin{pmatrix}
     g_L^2 (v_{dn}^2 + v_{un}^2)\ &~~~ -2 g_L g_R v_{dn} v_{un} \\[4pt]
     -2 g_L g_R v_{dn} v_{un}\ &~~~ g_R^2 (v_{dn}^2 + v_{un}^2 + 2\ V_\nu^2 ) 
    \end{pmatrix} \, .
    \label{eq:WLR}
\end{equation}
The mixing angle between $W_L^\pm$ and $W_R^\pm$ is strongly constrained to be $\leq 4 \times 10^{-3}$ from strangeness changing nonleptonic decays of hadrons \cite{Donoghue:1982mx}, as well as from $b\to s \gamma$ decay \cite{Babu:1993hx}, independent of the mass of the heavier $W_R^\pm$ state. LHC experiments have set a lower limit of about 4 TeV on $W_R^\pm$ mass from its production and subsequent decays into dijets, which translates into a lower limit of $V_\nu> 9$ TeV within the model, which follows from Eq. (\ref{eq:WLR}).  

The $T$-odd charged gauge bosons mass matrix is given, in the basis $(V_L^{\mu+} , V_R^{\mu+})$,  as
\begin{equation}
    M_{V_{L,R}^+}^{2} = \frac{1}{2}
    \begin{pmatrix}
     g_L^2 (v_{un}^2 + V_{n}^2 + 2\ (V_N^2 + V_\nu^2) )\ &~~~ -2 g_L g_R v_{un} V_{n} \\[4pt]
     -2 g_L g_R v_{un} V_{n}\ &~~~ g_R^2 (v_{un}^2 + V_{n}^2 + 2\ V_N^2 )
    \end{pmatrix} \, .
\end{equation}

In the neutral gauge boson sector, the $T$-odd $V^{\mu0}_{L,R}$ fields decouple from the $T$-even neutral gauge bosons, with no mixing among its real and imaginary components defined in Eq.~\eqref{eq:gaugeV}. The mass matrix for the real components in the basis $(W_{6L}^\mu, W_{6R}^\mu)$ reads as
\begin{equation}
    M_{W_{6LR}}^2 = \frac{1}{2} 
        \begin{pmatrix}
         g_L^2 (v_{dn}^2 + V_n^2 + 2\ (V_N^2 + V_\nu^2))\ &~~~ -2 g_L g_R v_{dn} V_n \\[4pt]
         -2 g_L g_R v_{dn} V_n\ &~~~ g_R^2 (v_{dn}^2 + V_n^2 + 2\ (V_N + V_\nu)^2)
        \end{pmatrix} \, .
\end{equation}
The mass matrix for the imaginary components in the basis $(W_{7L}^\mu, W_{7R}^\mu)$ reads as
\begin{equation}
    M_{W_{7LR}}^2 = \frac{1}{2} 
        \begin{pmatrix}
         g_L^2 (v_{dn}^2 + V_n^2 + 2\ (V_N^2 + V_\nu^2))\ &~~~ -2 g_L g_R v_{dn} V_n \\[4pt]
         -2 g_L g_R v_{dn} V_n\ &~~~ g_R^2 (v_{dn}^2 + V_n^2 + 2\ (V_N - V_\nu)^2)
        \end{pmatrix} \, .
        \label{eq:W7LR} 
\end{equation}
As can be seen from these matrices, one of the eigenstates, identified  approximately as $W_{7R}^\mu$, can be relatively light compared to all other gauge bosons. This is realized in the limit $V_N \simeq V_\nu \gg V_n$.  This is the vector boson dark matter candidate in the model. We define a parameter $a$ as
\begin{equation}
    (V_N - V_\nu) = a V_n \, 
    \label{eq:lightcond}
\end{equation}
with $a$ being a dimensionless parameter less than or of order unity, 
which we shall see is consistent with the minimization of the Higgs potential (cf. Eqs.~\eqref{eq:C1}-\eqref{eq:C3}- below).  Choosing $V_N = 9$ TeV, which is the lowest allowed value in this scheme from $W_R$ mass limits, the masses of $(W_R^{\mu+}, V_L^{\mu+}, V_R^{\mu+}, W_{6L}^\mu, W_{6R}^\mu, W_{7L}^\mu)$ are respectively $(3.96, 8.0, 3.96, 8.0, 7.92, 8.0)$ TeV, with $W_{7R}^\mu$ being much lighter with its mass given by $g_R V_n \sqrt{a^2+1/2}$. We shall see from the dark matter relic density constraints that this mass should lie below 900 GeV, which can be realized by choosing $V_n \sim $ TeV, and $a$ less than or of order unity.


In the $T$-even neutral gauge boson sector there are mixings among $(W_{3L}^\mu, W_{3R}^\mu , W_{8L}^\mu, W_{8R}^\mu)$  to produce the physical gauge bososn $A^\mu$, $Z^{\mu}_L$, $Z_{1 R}^\mu$, and $Z_{2 R}^\mu$ states,  We adopt the following convenient field definitions:
\begin{align}
    A^\mu &= \frac{\sqrt{3} g_R\ W_{3L}^\mu + \sqrt{3} g_L\ W_{3R}^\mu + g_R\ W_{8L}^\mu + g_{L}\ W_{8R}^\mu }{2 \sqrt{g_L^{2}+g_R^{2}}} \, , \\
    Z^{\mu}_L &= \frac{(g_R^2 + 4 g_L^2)\ W_{3L}^\mu - 3 g_L g_R\ W_{3R}^\mu - \sqrt{3} g_R^2\ W_{8L}^\mu - \sqrt{3} g_L g_R\ W_{8R}^\mu }{2 \sqrt{(g_L^2 + g_R^2) (4 g_L^2 + g_R^2)}} \, ,\\
    Z_{1R}^\mu &= \frac{-g_L\ W_{8L}^\mu + g_R\ W_{8R}^\mu}{\sqrt{g_L^2 + g_R^2}} \, , \\
    Z_{2R}^\mu &= \frac{-(g_L^2 + g_R^2)\ W_{3R}^\mu + \sqrt{3} g_L g_R\ W_{8L}^\mu + \sqrt{3} g_L^2\ W_{8R}^\mu}{\sqrt{(g_L^2 + g_R^2) (4 g_L^2 + g_R^2)}}  \, .
\end{align}
In this basis, the massless photon field $A^\mu$  decouples from the rest, leading to a $3\times 3$ mass matrix defined as
\begin{equation}
    (M_{Z_{LR}}^2)_{ij} = m_{ij} \, ,
\end{equation}
with the elements $m_{ij}$ given by
\begin{align}
    m_{11} &= \frac{2 g_L^2 (g_L^2 + g_R^2) (v_{dn}^2 + v_{un}^2)}{4 g_L^2 + g_R^2} \, , \\
    m_{12} &= \frac{g_L (g_L^2 + g_R^2) (v_{dn}^2- v_{un}^2)}{\sqrt{3\ (4 g_L^2 + g_R^2)}} \, ,\\
    m_{13} &= \frac{g_L g_R (g_L^2 + g_R^2) (v_{dn}^2 + v_{un}^2)}{4 g_L^2 + g_R^2} \\
    m_{22} &= \frac{1}{6 (g_L^2 + g_R^2)} \big[ (g_L^2 +g_R^2)^2 (v_{dn}^2 + v_{un}^2 + 4 (4 V_N^2 + V_n^2)) + 4 (g_R^2 - 2 g_L^2)^2 V_\nu^2 \big] \, ,\\
    m_{23} &= \frac{g_R (g_L^2 + g_R^2)^2 (v_{dn}^2 - v_{un}^2) + 4 g_R (-8 g_L^4 + 2 g_L^2 g_R^2 + g_R^4) V_\nu^2}{2 \sqrt{3} (g_L^2 + g_R^2) \sqrt{4 g_L^2 + g_R^2}} \, ,\\
    m_{33} &= \, \frac{g_R^2 (g_L^2 + g_R^2) (v_{dn}^2 + v_{un}^2)}{2 (4 g_L^2 + g_R^2)} + \frac{2 g_R^2 (4 g_L^2 + g_R^2) V_\nu^2}{g_L^2 + g_R^2} .
\end{align}
In the limit of unbroken electroweak symmetry, the masses for the heavy neutral fields are given by
\begin{align}
    M_{Z'_1, Z'_2}^2 = \frac{1}{3} \Big( (g_L^2 + g_R^2) (V_n^2 + 4 (V_N^2 + V_\nu^2)) \mp \sqrt{\Delta} \Big) \, ,
\end{align}
where $\Delta = (g_L^2 + g_R^2)^2 \big[(4 V_N^2 + V_n^2)^2 + 16 V_\nu^4 \big]+ 4 (2 g_L^2 - 8 g_L^2 g_R^2 - g_R^4) (4 V_N^2 + V_n^2) V_\nu^2 $. In the limit where $V_N-V_\nu \ll V_N$ as defined in Eq.~\eqref{eq:lightcond} and to the zeroth order in $V_n$ we have
\begin{equation}
     M_{Z'_1, Z'_2}^2 = \frac{4}{3}\ V_N^2\ \left[ 2 (g_L^2 + g_R^2) \mp |-2 g_L^2 + g_R^2| \right]~.
\end{equation}
If we set here the experimental lower limit of about 5 TeV on the $Z'$ mass from LHC searches, we would find $V_N \geq 5.7$ TeV, which is less stringent than the limit $V_N \ge1 9$ TeV  obtained from the $W_R$ searches at the LHC.  The masses of the gauge bosons $(Z'_1, Z'_2)$ when $V_N = V_\nu =9$ TeV are found to be $(7.92, 13.8)$ TeV.



\subsection{Higgs potential analysis}\label{sec:higgs}
\normalsize 

Here we construct and analyze the Higgs potential involving one copy of $(1,3,3^*)$, denoted by $\Phi_i^\alpha$ with $i$ and $\alpha$ being the $SU(3)_L$ and $SU(3)_R$ indices and one copy of $(1,6,6^*)$ field, denoted as  $\chi_{ij}^{\alpha\beta}$. The field $\chi$ obeys the symmetry properties $\chi_{ij}^{\alpha\beta} = \chi_{ji}^{\alpha\beta} = \chi_{ij}^{\beta \alpha}$.  While the full theory contains two additional copies of $\Phi(1,3,3^*)$ fields, here we note that only one combination of the three, defined as $(V_1 \Phi_1 + V_2 \Phi_2 + V_3 \Phi_3)/\sqrt{V_1^2+V_2^2+V_3^2}$, acquires a VEV along the SM singlet direction, while the two orthogonal combinations to this state only acquire electroweak scale VEVs.  It is this field with the VEV along the SM singlet direction that we keep in our analysis, which we denote simply as $\Phi_{i}^\alpha$. This should be an excellent approximation, as we ignore electroweak symmetry breaking effects in our analysis.  Certain scalar fields will be found to have masses below $V_N \simeq V_\nu$, which would be also maintained even with the inclusion of two additional $\Phi$ fields.  The terms involving these additional $\Phi$ fields can mix the fields we keep, which would only lower the masses of the unmixed states further.

The most general renormalizable potential for ($\chi, \Phi$) can be written as
\begin{equation}
    V = V(\chi) + V(\phi) + V(\Phi,\chi) \, ,
    \label{eq:pot}
\end{equation}
where
\begin{eqnarray}
V(\chi) &=& -m_\chi^2\, \chi_{ij}^{\alpha\beta} \chi^{ij}_{\alpha\beta}  + \{\mu_\chi\, \chi_{ij}^{\alpha\beta} \chi_{kl}^{\gamma\delta} \chi_{mn}^{\mu\nu} \epsilon^{ikm} \epsilon^{jln} \epsilon_{\alpha\gamma\mu} \epsilon_{\beta\delta\nu} +h.c.\} \nonumber \\
&+&\lambda_1\, \chi_{ij}^{\alpha\beta} \chi^{ij}_{\alpha\beta} \chi_{kl}^{\gamma\delta} \chi^{kl}_{\gamma\delta} + \lambda_2\, \chi_{ij}^{\alpha\beta} \chi^{ij}_{\alpha\gamma} \chi_{kl}^{\gamma\delta} \chi_{\beta\delta}^{kl} + \lambda_3\, \chi_{ij}^{\alpha\beta} \chi^{ij}_{\gamma\delta} \chi_{kl}^{\gamma\delta} \chi^{kl}_{\alpha\beta} \nonumber \\
&+& \lambda_4 \,\chi_{ij}^{\alpha\beta} \chi_{\alpha\beta}^{jk} \chi_{kl}^{\gamma\delta}\chi_{\gamma\delta}^{li} + \lambda_5\, \chi_{ij}^{\alpha\beta} \chi_{\alpha\gamma}^{jk} \chi_{kl}^{\gamma\delta}\chi^{li}_{\beta\delta}~.
\end{eqnarray}
Here we have used the notation $\chi^{ij}_{\alpha\beta} = (\chi_{ij}^{\alpha\beta})^*$. 
The remaining parts of the potential are given by\footnote{Symmetry breaking by a single $\Phi$ field has been analyzed in Ref. \cite{Bai:2017zhj}.}
\begin{equation}
    V(\Phi) = -m_\Phi^2\ \Phi_i^\alpha \Phi_\alpha^i + \{ \mu_{\Phi}\ \Phi_i^\alpha \Phi_j^\beta \Phi_k^\gamma \epsilon^{ijk} \epsilon_{\alpha \beta \gamma} + h.c. \} + \lambda_6\ \Phi_i^\alpha \Phi_\alpha^i \Phi_j^\beta \Phi_\beta^j + \lambda_7\ \Phi_i^\alpha \Phi_\alpha^j \Phi_j^\beta \Phi_\beta^i \, ,
\end{equation}
and
\begin{align}
    V(\Phi, \chi) &= \{ \mu_{\phi \chi}\ \Phi_i^\alpha \Phi_j^\beta \chi_{\alpha \beta}^{i j} + h.c. \} + \lambda_8\ \Phi_i^\alpha \Phi_\alpha^i \chi_{jk}^{\beta \gamma} \chi_{\beta \gamma}^{jk} + \lambda_9\ \Phi_i^\alpha \Phi_\beta^i \chi_{kl}^{\beta \gamma} \chi_{\alpha \gamma}^{k l} \nonumber \\
    &+ \lambda_{10}\ \Phi_i^\alpha \Phi_\alpha^j \chi_{jl}^{\beta \gamma} \chi_{\beta \gamma}^{il} + \lambda_{11}\ \Phi_i^\alpha \Phi_\beta^j \chi_{jl}^{\beta \gamma} \chi_{\alpha \gamma}^{i l} + \Big\{ \lambda_{12}\ \Phi_i^\alpha \chi_{\beta \gamma}^{jk} \chi_{l j}^{\beta \delta} \chi_{mk}^{\kappa \gamma} \epsilon^{ilm} \epsilon_{\alpha \delta \kappa} \nonumber \\ 
    &+ \lambda_{13}\ \Phi_i^\alpha \Phi_j^\beta \Phi_\gamma^k \chi_{kl}^{\delta \gamma} \epsilon^{ijl} \epsilon_{\alpha \beta \delta} + \lambda_{14}\ \Phi_i^\alpha \Phi_j^\beta \chi_{kl}^{\gamma \delta} \chi_{mn}^{\rho \sigma} \epsilon^{ikm} \epsilon^{jln} \epsilon_{\alpha \gamma \rho} \epsilon_{\beta \delta \sigma} + h.c. \Big\}  \, .
\end{align}
Note that the VEVs of  Eq.~\eqref{eq:vevs} can in general be complex.  Here for simplicity we take them to be real.  This is achieved by assuming that the complex parameters in the Higgs potential are all real.
 By inserting the VEV of Eq.~\eqref{eq:vevs} in Eq.~\eqref{eq:pot}, and by minimizing the potential, mass matrices  for the charged and neutral Higgs bosons can be constructed. A complete analysis of the spectrum that includes  electroweak symmetry breaking effects is beyond the scope of the present work. However, in the electroweak symmetric limit, the charged, scalar, and pseudoscalar mass matrices in the $T$-even and $T$-odd sectors of the theory can be built from the Higgs potential of Eq.~\eqref{eq:pot}. We show the consistency of symmetry breaking and identify the light scalars that survive to scales below $V_N \simeq V_\nu$.
 
By inserting the VEV $\langle \Phi_3^3 \rangle = V_n$, $\langle \chi_{33}^{22} \rangle = V_\nu$, and $\langle \chi_{33}^{33} \rangle = V_N$, we obtain the following condition for the potential to be extremum:
\begin{align}
    \mu_{\phi \chi} V_n^2 + V_N \big[-m_\chi^2 + 2 \lambda V_N^2 + \lambda' V_n^2  + 2(\lambda_1 + \lambda_3 + \lambda_4) V_\nu^2 \big] &= 0  \label{eq:C1} \, ,\\
  V_\nu \big[-m_\chi^2 + (\lambda_{10} + \lambda_8) V_n^2 + 2 (\lambda_1 + \lambda_3 + \lambda_4) V_N^2 + 2 \lambda V_\nu^2 \big]    &= 0 \label{eq:C2} \, ,\\
  V_n \big[-m_\Phi^2  + 2 \mu_{\phi \chi} V_N + \lambda' V_N^2 + (\lambda_{10} + \lambda_{8}) V_\nu^2 + 2 (\lambda_6 + \lambda_7) V_n^2 \big]  &= 0 \, ,
   \label{eq:C3}
\end{align}
where we have defined $\lambda = \lambda_1 + \lambda_2 + \lambda_3 + \lambda_4 + \lambda_5$ and $\lambda' = \lambda_{8} + \lambda_{9} + \lambda_{10} + \lambda_{11}$. We eliminate $m_\chi^2, m_\Phi^2$, and $\mu_{\phi \chi}$ using these three conditions and keep the VEVs $V_N, V_\nu$ and $V_n$ as independent parameters. As noted earlier, none of these VEVs can be set to zero for consistency with fermion and gauge boson masses. The full spectrum for the masses of the Higgs fields are given in detail in Appendix \ref{sec:append}, where  SM quantum number each component has been properly identified. We have also identified the 12 Goldstone modes associated with spontaneous breaking of $SU(3)_L \times SU(3)_R$ down to $SU(2)_L \times U(1)_Y$.

\subsubsection{Necessary conditions for boundedness of the potential}\label{sec:bounded}
\begin{table}[]
\scriptsize
    \centering
    \begin{tabular}{|c|c|c|c|c|c|c|c|c|c|c|c|c|}
    \hline 
        {\bf Parameter} &  $\lambda_1$ & $\lambda_2$& $\lambda_3$&$\lambda_4$ &$\lambda_5$ & $\lambda_6$& $\lambda_7$& $\lambda_8$& $\lambda_9$& $\lambda_{10}$& $\lambda_{11}$ \\[3pt]
    \hline
        {\textbf{BP-I}} & 2.42 & -0.0061 & -0.49 & -1.7 & -0.0053 & 1.77 & 1.68 & 1.87 & 1.89 & -1.71 & 1.76    \\[3pt]
    \cline{2-12}
    \cline{2-12}
     & $\lambda_{12}$ & $\lambda_{13}$ & $\lambda_{14}$ & $a$  & $\mu_\chi$ & $\mu_\Phi$ & $V_N$ & $M_{DM}$ & $M_{h'}$  & $M_{H'}$ & $M_{\delta^{++}}$   \\[3pt] 
    \cline{2-12}
    & -1.21 & -0.64 & -1.06 & 0.68 & 0.18 & -0.74 & -0.16  & 0.076 & 0.057  & 0.165 & 0.133   \\[3pt]
    \cline{1-12}
    \end{tabular}
    
    \vspace{5mm}
    \begin{tabular}{|c|c|c|c|c|c|c|c|c|c|c|c|c|}
    \hline
    {\bf Parameter} & $\lambda_1$ & $\lambda_2$& $\lambda_3$&$\lambda_4$ &$\lambda_5$ & $\lambda_6$& $\lambda_7$& $\lambda_8$& $\lambda_9$& $\lambda_{10}$& $\lambda_{11}$ \\[3pt]
    \hline
        {\bf BP-II} & 1.96 & -0.0061 & -0.18 & -1.42 & -0.0025 & 0.54 & 0.74 & 1.10 & 1.89 & -1.14 & 1.81  \\[3pt]
    \cline{2-12}
    \cline{2-12}
     & $\lambda_{12}$& $\lambda_{13}$ & $\lambda_{14}$ & $a$  & $V_n$ & $\mu_\chi$ & $\mu_\Phi$  & $M_{DM}$ & $M_{h'}$ & $M_{H'}$ & $M_{\delta^{++}}$  \\[3pt]
    \cline{2-12}
     & -1.19 & -0.72 & -1.03 & 0.61 & 0.17 & -0.32 & -0.41  & 0.069 & 0.095 & 0.187 & 0.118  \\[3pt]
    \cline{1-12}
    \end{tabular}
    \caption{Values of the parameters that satisfies the necessary boundedness conditions defined in Eqs.~\eqref{eq:bound1}-\eqref{eq:bound2} and positivity of the square of mass matrices given in Appendix \ref{sec:append}. Here all the mass parameters are in units of the larger VEV $V_N \simeq V_\nu$. Masses of the remaining scalar fields are of order one (in units of $V_N$) and heavier than $M_{H'}$. Their numerical values are given in Appendix \ref{sec:append}. The state $\tilde{h}$ is somewhat light and has an intermediate scale mass given by $M_{\tilde{h}} = 0.65\ (0.39)$ for the {\tt BP-I} ({\tt BP-II}).  }
    \label{tab:bound}
\end{table}
The full set of necessary and sufficient conditions on the quartic couplings of the potential to be bounded from below is not easily tractable in the model. However, certain necessary conditions of phenomenological interest can be analyzed analytically. We focus only on the neutral singlet fields $\Phi_3^3$, $\chi_{33}^{22}$, and $\chi_{33}^{33}$. The quartic terms $V^{(4)} (\Phi, \chi)$ form a vector space spanned by the real-valued vectors $x^T = \{V_n^2, V_N^2, V_\nu^2 \}$ which can be written as
\begin{equation}
    V^{(4)} = \frac{1}{2} x^T \hat{\lambda}\ x \, ,
\end{equation}
where
\begin{equation}
  \hat{\lambda} = \begin{pmatrix}
    \lambda_6 + \lambda_7 &~~~ \lambda'/2 &~~~ (\lambda_{10} + \lambda_8)/2 \\
   \lambda'/2 &~~~ \lambda &~~~ ( \lambda_1 +  \lambda_3 + \lambda_4) \\
   (\lambda_{10} + \lambda_8)/2 &~~~ ( \lambda_1 +  \lambda_3 + \lambda_4) &~~~ \lambda
    \end{pmatrix}~.
    \label{eq:bound}
\end{equation}
Here $\lambda = \lambda_1 + \lambda_2 + \lambda_3 + \lambda_4 + \lambda_5$ and $\lambda' = \lambda_{8} + \lambda_{9} + \lambda_{10} + \lambda_{11}$, as defined earlier. The necessary and sufficient conditions for the boundedness of this restricted potential can be derived from the co-positivity of the real symmetric matrix $\hat{\lambda}$ of Eq.~\eqref{eq:bound}, which are \cite{HADELER198379, Klimenko:1984qx}:
\begin{align}
    &\lambda_6 + \lambda_7 \geq 0 \, , \\
    \label{eq:bound1}
    &\lambda \geq 0 \, , \\
    & \lambda'/2 \geq - \sqrt{(\lambda_6 + \lambda_7) \lambda} \, , \\
    & (\lambda_{10}+ \lambda_8)/2 \geq - \sqrt{(\lambda_6 + \lambda_7) \lambda} \, , \\
    & (\lambda_{1}+ \lambda_3 + \lambda_4) \geq - \lambda \, , \\
    &\lambda'/2 \sqrt{\lambda} + (\lambda_1 + \lambda_3 + \lambda_4) \sqrt{\lambda_6+ \lambda_7} \nonumber\\
    &~~~~~~~~~+ 1/2(\lambda_8+ \lambda_{10}) \sqrt{\lambda} + \sqrt{(\lambda_6+\lambda_7)\lambda^2} \geq 0\ \text{or}\ \text{det} \hat{\lambda} \geq 0 \, .
    \label{eq:bound2}
\end{align}
In Table~\ref{tab:bound}, we show two benchmark points that satisfies the necessary boundedness conditions defined in Eqs.~\eqref{eq:bound1}-\eqref{eq:bound2}. The same parameter set also satisfies the positivity criteria of the mass eigenvalues of all scalar fields given in Appendix \ref{sec:append}. We have also shown the masses of the lighter states in the theory in Table~\ref{tab:bound}.  All states not shown in the table have masses much larger, and we show their numerical values for the two benchmark points 
 in Appendix \ref{sec:append}.

\subsubsection{Light scalars in the model}

In this subsection we discuss the light scalar fields that are present in the model. We focus on the hierarchy $V_N \simeq V_\nu \gg V_n$, which generates a light vector boson DM candidate naturally.  
The light Higgs states in the model, in this approximation, are found to be
\begin{equation}
    \{\delta^{++}, \,h', \,H', \,\tilde{h} \}
    \label{fields}
\end{equation}
which have masses of order $V_n$,  the lower scale of $O(2)_R$ symmetry breaking. Here the $h'$ field can be identified as the remnant of this $O(2)_R$ breaking, while the others remain light owing to the constraints arising from the Higgs potential.  $H'$ here is a $T$-odd field, which makes it a possible  candidate for scalar DM in the theory. The other fields in Eq. (\ref{fields}) are all $T$-even scalars. 

The reason for the lightness of these scalar fields can be identified as follows.  With only the $\chi$ field participating in symmetry breaking, the extremum $V_N = V_\nu$ (which we pursue) is a saddle point, and not a minimum.  If we ignore the contribution to the masses from the $\Phi$ field, certain squared masses from $\chi$ would be positive, while certain others would be negative.  It turns out that the potential with only the $\chi$ field prefers to break the $SU(3)_L \times SU(3)_R$ symmetry down to $SU(2)_L \times SU(2)_R \times U(1)_Y$ by setting either $V_\nu =0$ or $V_N = 0$. With the inclusion of the $\Phi$ field, the situation changes. However, turning the would-be negative mass-squared states into positive mass-squared states from $\chi$ would require certain quartic couplings, viz., $\lambda_2+\lambda_5$, to be of order $V_n^2/V_N^2$. This in turn causes the states listed in Eq. (\ref{fields}) to remain light with masses of order the smaller VEV $V_n$.

As can be seen from Eq.~\eqref{eq:dobulecharged}, positivity of the squared mass of $\delta^{++}$ demands $\lambda_2 + \lambda_5 < 0$. However, the positivity condition  of the squared mass of the state given in Eq.~\eqref{eq:Msig3} would require $\lambda_2 + \lambda_5 > 0$, if we take the approximation $V_N = V_\nu$ and $V_n =0$.  This contradiction can be avoided by making use of the condition
 of Eq.~\eqref{eq:lightcond} if we define $\lambda_2 + \lambda_5 = b\ (\lambda_9 + \lambda_{11})\ V_n^2/V_N^2$, where $b$ is of order unity.  
Thus, the masses of the doubly charged scalar $\delta^{++}$ of Eq. (\ref{eq:dobulecharged}) and the $T$-odd singlet scalar $H'$ given in Eq.~\eqref{eq:Msig3} become
\begin{eqnarray}
    M_{\delta^{++}} &=& \sqrt{(\lambda_9 + \lambda_{11})\ b}\ V_n \\
    M_{H'} &=& \frac{1}{\sqrt{2}}\ \sqrt{ (1 + 2 a^2)\ (1 + 8 b)\ (\lambda_{11} + \lambda_{9})}\ V_n \, .
\end{eqnarray}
Thus these two states are lighter than the largest VEV in the theory $V_N$. 

 There are two more $T$-even weak-singlet neutral scalar fields that are light in the theory. These are linear combinations of the fields $(\text{Re}[\chi_{33}^{22}], \text{Re}[\chi_{33}^{33}], \text{Re}[\Phi_{3}^{3}])$, and the mass-squared terms are given in Eqs.~\eqref{eq:lightH1}-\eqref{eq:lightH2}. We work in the limit of $V_n/V_N \ll 1$ and make a $45^\circ$ rotation among the first two states which decouples the heavier state. The resulting mass matrix reads as
\begin{equation}
    M^2 = \begin{pmatrix}
     8 \hat{\lambda}\ V_\nu (V_\nu + a V_n ) &~~ 0 &~~~ 0 \\
     0 &~~ \frac{1}{2} (1+8b) \tilde{\lambda}\ V_n^2 &~~~ - 4 \sqrt{2}\ a b\ \tilde{\lambda}\ V_n^2 \\
     0 &~~ - 4 \sqrt{2} a b\ \tilde{\lambda}\ V_n^2 &~~~ \Big[ 4 (\lambda_6 + \lambda_7) - \frac{(\lambda_{10}+ \lambda_8)^2}{\hat{\lambda}} \Big] V_n^2
    \end{pmatrix} \, ,
    \label{eq:lighteven}
\end{equation}
where $\tilde{\lambda} = \lambda_{11} + \lambda_{9}$ and $\hat{\lambda} = \lambda_1 + \lambda_3 + \lambda_4$. It is clear from the  Eq.~\eqref{eq:lighteven} that two of the states have masses proportional to $V_n$ which will remain light. In our numerical study of DM annihilation, for concreteness we work in the limit of $ab \ll 1$ so that these two light states become pure states. We denote the second state as $\tilde{h}$ and the third state as $h'$. Note that if $h'$ and $\tilde{h}$ are lighter than vector boson DM, it  can annihilate into pairs of these states.  

We briefly comment on the possibility of scalar DM within the theory.  The state $H'$ is a SM singlet and $T$-odd. If it is the lightest $T$-odd particle, it would constitute the DM of the universe. Unlike the scalar singlet DM model within the context of SM \cite{Silveira:1985rk,McDonald:1993ex}, here $H'$ can annihilate into pairs of $h'$ and $\tilde{h}$, provided that these $T$-even scalars are lighter than $H'$. This would allow for a significantly larger region of  parameter space, analogous to complex singlet DM \cite{Gonderinger:2012rd,Coito:2021fgo}. 

\section{Collider Implications}\label{sec:LHC}
The model developed, with TeV scale trinification symmetry can be readily tested at the Large Hadron Collider (LHC). The existence of vector-like quarks and  leptons, in addition to the vector boson dark matter candidate, results in a rich phenomenology for the LHC. In this section, we explore the potential smoking gun signals associated with vector boson dark matter and current LHC constraints.

\begin{figure}[htb!]
    \centering
    \includegraphics[width=0.3\textwidth]{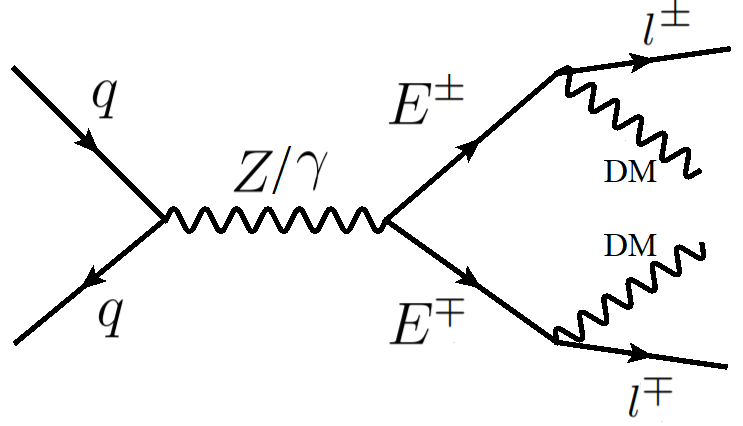}
    \caption{Representative Feynman diagrams for vector-like lepton production and subsequent decays at the LHC.
    }
    \label{fig:col1}
\end{figure}

\begin{figure}[htb!]
    \centering
    \includegraphics[width=0.6\textwidth]{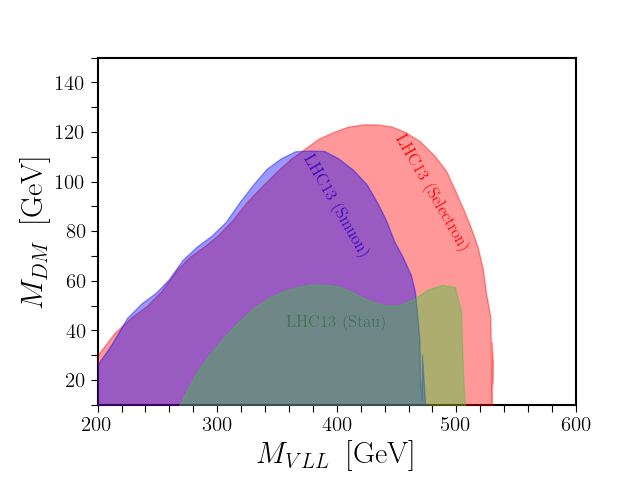}
    \caption{Current LHC limits on vector like leptons in our model. The shaded regions are excluded by LHC searches of $pp \to l^+ l^- +{E\!\!\!\!/}_{T}$ with $l = e,\,\mu,\,\tau$.}
    \label{fig:col2}
\end{figure}
Vector-like leptons will be dominantly pair-produced at the LHC through the $s-$ channel $Z/\gamma$ exchange (see Fig.~\ref{fig:col1}), and then decay to leptons and vector boson dark matter, yielding a promising $pp \to l^+ l^- +{E\!\!\!\!/}_{T}$ signature. This will resemble the signals of supersymmetric slepton searches \cite{CMS:2018eqb, ATLAS:2019ucg} leading to dilepton and missing energy signatures. It is worth noting that the pair-production rate of vector-like leptons (fermion) in our scenario is substantially larger than the pair-production rate of slepton (scalar). Non-observation of any new physics signal in the $pp \to l^+ l^- +{E\!\!\!\!/}_{T}$ channel \cite{CMS:2018eqb, ATLAS:2019ucg} imposes tight bounds on the sub-TeV vector-like lepton masses. We recast the bounds on the vector-like leptons in our model using the most current selectron \cite{ATLAS:2019ucg}, smuon \cite{ATLAS:2019ucg}, and stau \cite{CMS:2018eqb} search limits. In order to do that, we implement our model file in the {\tt FeynRules} package~\cite{Christensen:2008py} and then compute the signal cross sections using {\tt MadGraph5aMC@NLO}~\cite{Alwall:2014hca}. Then, assuming that the cut efficiencies are identical in both scenarios, we compare the signal cross-sections to the experimental upper bounds. We find that the selectron search \cite{ATLAS:2019ucg} yields the most stringent constraint on the vector-like lepton mass, with the exclusion region reaching $\sim$ 630 GeV. We summarize these current LHC limits on the vector-like leptons in Fig.~\ref{fig:col2}.

\begin{figure}[htb!]
    \centering
    \includegraphics[width=0.99\textwidth]{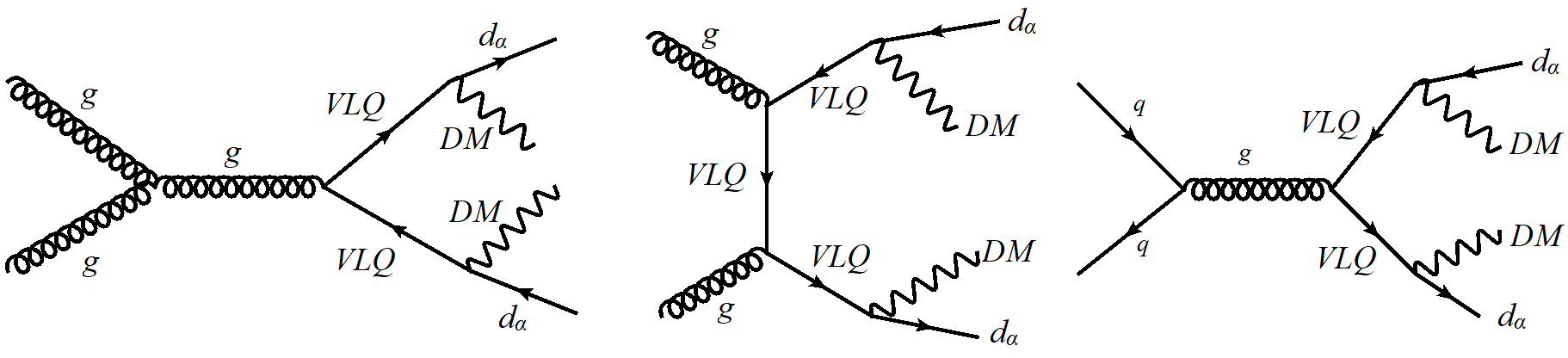}
    \caption{Leading Feynman diagrams for vector-like quark production and subsequent decays at the LHC.
    }
    \label{fig:col3}
\end{figure}

\begin{figure}[htb!]
    \centering
    \includegraphics[width=0.6\textwidth]{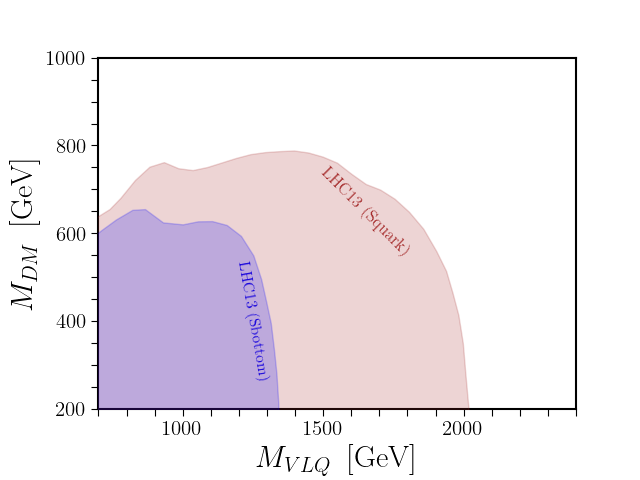}
    \caption{Current LHC limits on vector-like quarks in the model. The shaded regions are excluded by $pp \to jj+ +{E\!\!\!\!/}_{T}$ searches. The inner shaded region corresponds to the jets identified as $b$-jets. }
    \label{fig:col4}
\end{figure}

Unlike vector-like lepton pair production, vector-like quark pair production at the LHC will be dominated by gluon-gluon fusion processes. Representative Feynman diagrams for the dominant production processes are shown in Fig.~\ref{fig:col3}.  The vector-like quark pair production rate is exclusively dictated by the strong coupling and vector-like quark masses. In our model, vector-like quarks decay to vector boson dark matter along  with down-type quarks. This will result in  $pp \to jj+ +{E\!\!\!\!/}_{T}$ and  $pp \to b\bar{b}+ +{E\!\!\!\!/}_{T}$  signatures.  In supersymmetry, squark or gluino can lead to the identical final state signatures, and there are specific experimental searches \cite{ATLAS:2020syg, CMS:2019zmd} looking at $pp \to jj+ +{E\!\!\!\!/}_{T}$ and  $pp \to b\bar{b}+ +{E\!\!\!\!/}_{T}$  signatures. It is worth noting that the rates of squark or gluino creation differ significantly from the rates of vector-like quark pair formation. We compute the production rate using {\tt MadGraph5aMC@NLO}~\cite{Alwall:2014hca} and recast the limit from supersymmetric squark searches \cite{ATLAS:2020syg} by comparing the signal cross-sections to the experimental upper bounds on the cross-sections assuming the cut efficiencies are identical in both scenarios. We find that the bound on vector-like quarks leading to the $pp \to jj+ +{E\!\!\!\!/}_{T}$ signature is $\sim$ 2 TeV, whereas the bound on vector-like quarks leading to the $pp \to b\bar{b}+ +{E\!\!\!\!/}_{T}$ signature is 1.3 TeV. All these limits on vector-like quarks are summarized in Fig.~\ref{fig:col4}. It should be noted that vector-like quarks can be created individually via quark-gluon fusion processes, resulting in a monojet signature in conjunction with missing energy. The single production rate of vector-like quarks is affected by both the gauge coupling $g_R$ and the strong coupling constant. We find that the single production rate is significantly suppressed, and the limit from the pair-production scenario is stronger than the single production case. It can be foreseen that the next collider experiments like the HL-LHC, HE-LHC, and FCC-hh will provide ample opportunity to test this vector-like fermion portal dark matter scenario in its complete range.

\section{Vector Boson Dark Matter Phenomenology}\label{sec:DMrelic}

In this section we analyze the allowed parameter space of the vector boson DM by numerically evaluating the relic abundance using the software MicrOmegas \cite{Belanger:2014vza, Belanger:2020gnr}. The model is implemented in CalcHEP \cite{Belyaev:2012qa} by using LanHEP \cite{Semenov:2014rea}. In the limit of $(V_N - V_\nu) = a V_n$, with $V_n \ll V_N$ and $a$ being of order one, the mass of dark matter $W_{7R}$ from Eq.~\eqref{eq:W7LR} reads as
\begin{equation}
    M_{W_{7R}} \equiv M_{DM} \simeq \frac{1}{\sqrt{2}} g_R V_n\ \sqrt{1+2 a^2} \, .
    \label{eq:DMmass}
\end{equation}
Moreover, since the DM mass and vector-like fermion masses are controlled by the same $O(2)_R$ breaking VEV $V_n$, one can obtain theoretical limit on the ratio of DM mass and vector-like fermion mass. Using Eq. (\ref{eq:DMmass}), together with Eq.~\eqref{eq:exoticquark} (and similarly Eq.~\eqref{eq:chargeL}) and demanding Yukawa $Y_q \leq 1.5$ (which is the perturbative unitarity limit on the quark Yukawa coupling), one obtains 
\begin{equation}
    M_{VLQ} \leq \frac{5 M_{DM}}{\sqrt{1+ a^2}} \, .
    \label{eq:VLQ}
\end{equation}
This sets a strong theoretical bound in the model parameters, making the model very predictive, as discussed below. It is important to stress that to avoid the DM decay into exotic plus a SM fermions, the mass of the DM should obey $M_{DM} < M_{VLQ},\, M_{VLL}$, where $VLQ$ ($VLL$) stands for vector-like quarks (leptons).

\subsection{Annihilation cross section and relic abundance}\label{sec:relic}
\begin{figure}
    \includegraphics[scale=0.35]{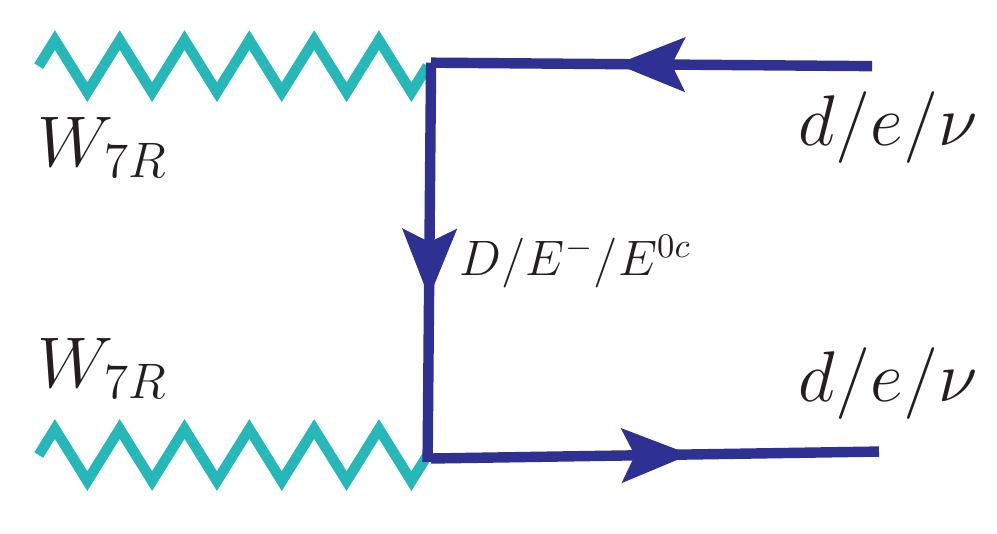} \hspace{3mm}
    \includegraphics[scale=0.35]{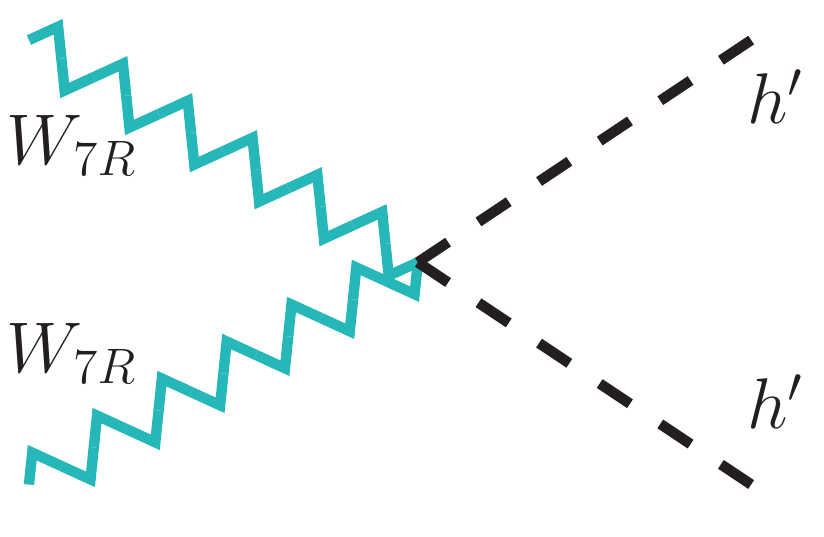} \hspace{3mm}
     \includegraphics[scale=0.35]{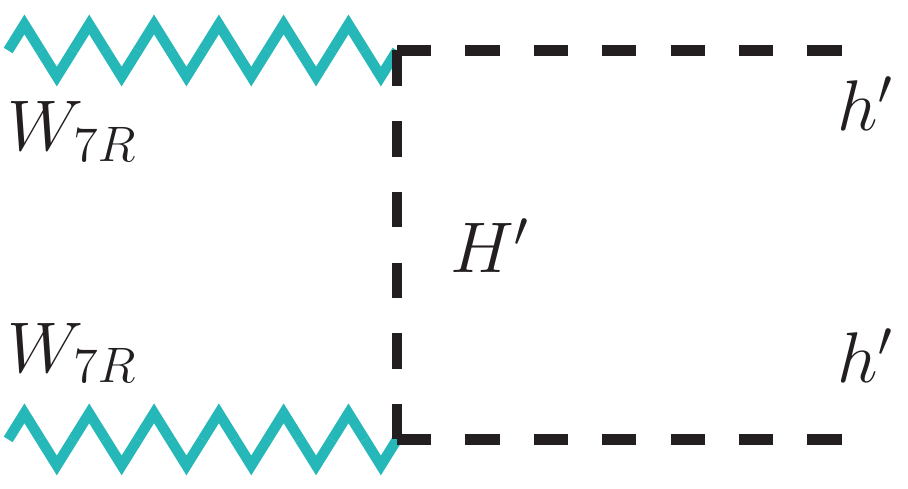} \hspace{3mm}
    \includegraphics[scale=0.2]{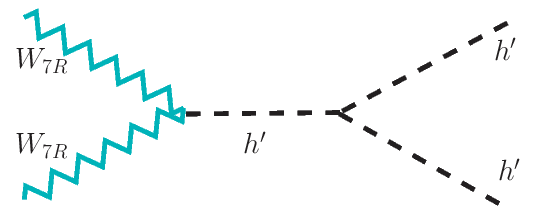}
    
    \vspace{3mm}
     ~~~~~~~~~~~(a)~~~\hspace{15mm}~~~~~~~~~~(b)~~~~~~~~~~~~~~~~~~~~~~~(c)~~~~~\hspace{20mm}~~~~~~~~~~(d)~~~~~~~~~
    \caption{Relevant Feynman diagrams that contribute to vector boson DM annihilation. }
    \label{fig:anni}
\end{figure}
\begin{figure}
    \centering
    \includegraphics[scale=0.25]{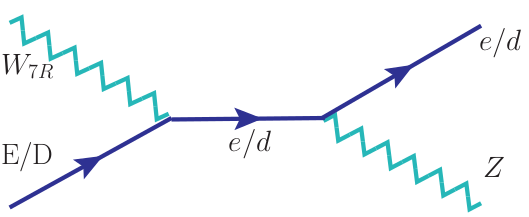}
    \caption{A typical Feynman diagram that contribute to co-annihilation of DM. There is an additional diagram with exotic charged lepton $E$ replaced by $E^{c0}$ and SM charged lepton replaced by neutrinos. }
    \label{fig:coanni}
\end{figure}
The relic density of DM is achieved though standard thermal freeze-out mechanism. Pairs of dark matter can annihilate through the three point and four point couplings  shown in Fig.~\ref{fig:anni} mediated by an off- or on-shell scalar bosons or vector-like fermions. The relevant gauge interactions of DM with the light degrees that are used for annihilation are given by
\begin{align}
 \mathcal{L} \supset  &  \frac{g_R}{2}\ W_{7R}^\mu\ \left[   i(E^- \bar{e}^- + \bar{\nu} E^{c0} + d_R \bar{D}_R + h.c. ) + (h' \partial_\mu H'- H' \partial_\mu h') \right]  \nonumber \\
 & + \frac{g_R^2}{8}  W_{7R}^\mu W_{7R \mu}\ (h' h' + H' H') \, .
 \label{eq:gaugeanni}
\end{align}
As already mentioned, the scalar fields that can be light in the framework are $T$-even neutral scalars $h'$  and $\tilde{h}$, a doubly charged scalar $\delta^{++}$, and $T$-odd neutral scalar $H'$, whose masses are given in Eq.~\eqref{eq:masshp}, Eq.~\eqref{eq:dobulecharged}, and Eq.~\eqref{eq:Msig3} respectively. For simplicity, we take masses of $\tilde{h}$, $H'$ and $\delta^{++}$ heavier than the vector boson  DM. Such a choice is consistent within our model as shown by the benchmark points in Table.~\ref{tab:bound}. With this simplification, a pair of DM can annihilate to (i) SM fermion pair ($d\bar{d}, e\bar{e}, \nu \bar{\nu}$) via the exchange of vector-like fermions ($D_R, E^-, E^{c0}$) through $t-$channel graph as shown in Fig.~\ref{fig:anni} (a), (ii) a pair of $h'$ via the exchange of $H'$, $h$ and through four point contact term as shown in Fig.~\ref{fig:anni} (b)-(d), and (iii)  pairs of SM particles through the mixing between $h'$ and SM Higgs $h$. The mixing between $h'$ and SM Higgs is heavily constrained by DM direct detection constraints \cite{Speller:2015mna, XENON:2018voc, LUX-ZEPLIN:2018poe, PandaX-4T:2021bab}. Here we turn off such couplings and avoid the constraint from direct detection. It is also important to note that if $T$ parity odd vector-like fermion mass is close to the DM mass, it can contribute to co-annihilation, which would  have strong impact on the relic abundance in the model. A typical graph for co-annihilation is shown in Fig.~\ref{fig:coanni}. 


\begin{figure}[!t]
    $\includegraphics[scale=0.48]{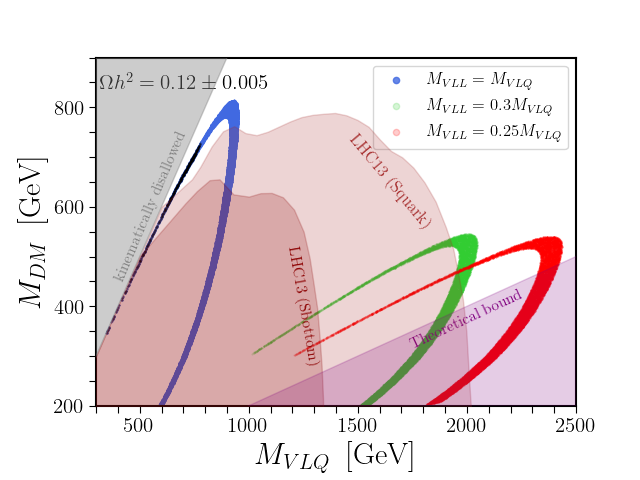}
    \includegraphics[scale=0.48]{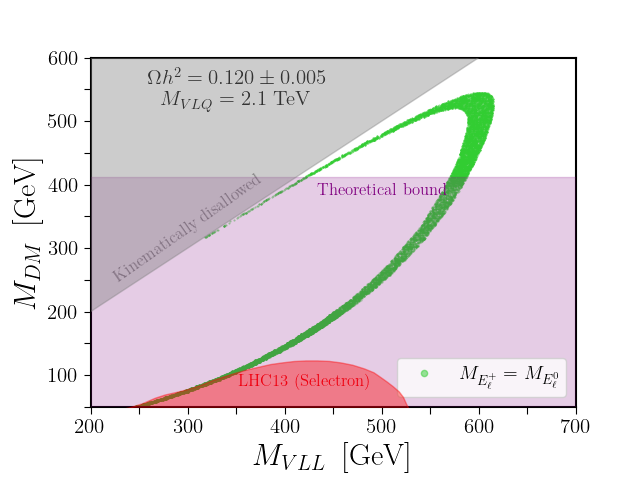}\\
    ~~\hspace{3.2cm}~~(a)~~~~~~~~~~~~~~~~~~\hspace{5cm}~(b)
    $
    \caption{Allowed regions in the mass of dark matter $M_{DM}$ and mass of vector-like fermion $M_{VLQ} (M_{VLL})$ plane that satisfy relic abundance requirement. The $T$-even $h'$ and $T$-odd $H'$ scalar masses are taken much heavier than the mass of vector-like fermions here, so that only $t$-channel process as shown in Fig.~\ref{fig:anni} (a) contributes. The gray and purple shaded regions are respectively from kinematically disallowed ($M_{DM} > M_{VLQ}$) region and theoretical bound ($M_{VLQ} > 5 M_{DM}$). In the left panel (blue, green, red) shades respectively represents allowed parameter space that satisfies the relic abundance for the choice  $M_{VLL}/M_{VLQ} = (1, 0.3, 0.25)$. Dark (light) brown corresponds to the exclusion region from $pp \to jj (bb) + {E\!\!\!\!/}_{T}$ obtained by recasting the supersymmetric squark (sbottom) searches. The black segment along the abundance curve is excluded from direct detection constraints \cite{XENON:2018voc}. In the right panel the green band corresponds to the allowed parameter space that satisfies the relic abundance for a fixed VLQ mass $2.1$ TeV. The red band is the exclusion region from $pp \to l^+ l^- +{E\!\!\!\!/}_{T}$ signature obtained by recasting supersymmetric slepton searches at the LHC.  }
    \label{fig:xx_D_sm}
\end{figure}

All the relevant couplings for the annihilation are determined  by the gauge coupling $g_R$ appearing in Eq.~\eqref{eq:gaugeanni}, except for the diagram in Fig.~\ref{fig:anni} (d), where the vertices are given by $\frac{g_R^2 V_n}{2\sqrt{2}} (W_{7R} W_{7R} h')$ and $\sqrt{2} \lambda_{h'} V_n (h'^3)$. The VEV $V_n$ can be traded off for the mass of DM given in Eq.~\eqref{eq:DMmass}, whereas the quartic coupling $\lambda_{h'}$ is given in Eq.~\eqref{eq:quartichp}. Thus the effective parameters that characterizes the relic abundance of DM in the model are:
\begin{equation}
    \{ M_{DM},\ M_{VLQ},\ M_{VLL},\ M_{h'} \} \, ,
\end{equation}
where we take the masses for all three fermion families to be equal. Note that the gauge coupling $g_R$ at energy scale 2 TeV is determined to be $g_R = 0.445$. The running of $g_R$ in the range $\mu = (0.5-5)$ TeV is not significant, and  does not effect the phenomenology by much. To explore the parameter space, we adopt two scenarios: (i) $M_{h'} > M_{DM}$, so that only the $t$-channel graph via the exchange of vector-like fermions in Fig.~\ref{fig:anni} (a) contributes to DM annihilation, and (ii) $M_{h'}$ fixed at a particular value such that when $M_{DM} > M_{h'}$ new annihilation modes open up as shown in Fig.~\ref{fig:anni} (b)-(d). It is important to point out that there are no allowed parameter space while only considering annihilation mode to a pair of $h'$, as one cannot take the mass of vector-like fermion much larger than the DM mass.  

\begin{figure}
    \centering
     \includegraphics[scale=0.6]{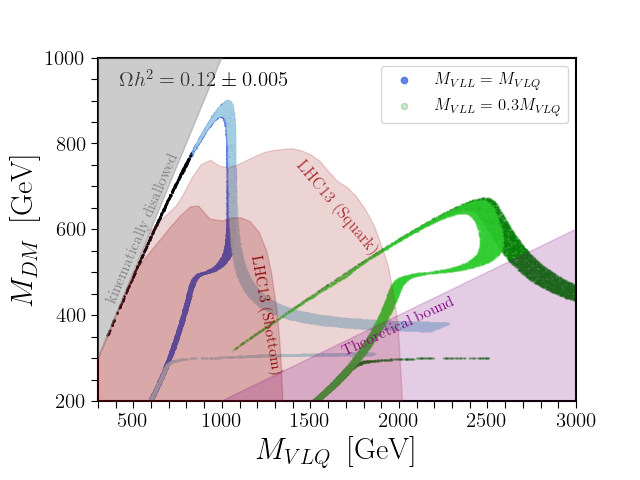}
    \caption{Allowed regions in the mass of dark matter $M_{DM}$ and mass of vector-like quark plane that satisfies relic abundance. The $T$-even $h'$ scalar mass is kept at fixed value such that when $M_{DM} > M_{h'}$, all the annihilation modes of in Fig.~\ref{fig:anni} are relevant. Light (dark) blue band corresponds to allowed parameter space with $M_{VLL}=M_{VLQ}$ that satisfies relic density for $M_{h'} = 300\ (500)$ GeV, whereas light (dark) green region corresponds to $M_{VLL}= 0.3\ M_{VLQ}$ for $M_{h'} = 300\ (500)$ GeV. The black segment along the abundance curve is excluded from direct detection constraints \cite{XENON:2018voc}.  }
    \label{fig:combined_plot}
\end{figure}

For the first case with $t$-channel annihilation of DM via vector-like fermions into SM fermions, the relevant parameters are $\{ M_{DM},\ M_{VLQ},\ M_{VLL} \}$. Here we explore two scenarios: (i) taking the ratio of VLQ and VLL fixed and varying mass of DM and VLQ, depicted in Fig~\ref{fig:xx_D_sm} (a), and (ii) fixing $M_{VLQ} = 2.1$ TeV so that the collider bounds are satisfied and varying in the plane of DM mass and VLL mass, which is depicted in Fig~\ref{fig:xx_D_sm} (b). The color shaded regions are  excluded regions in Fig~\ref{fig:xx_D_sm}. The gray and purple shaded regions are excluded respectively from kinematics ($M_{DM} > M_{VLQ}$) and theoretical bound ($M_{VLQ} > 5 \,M_{DM}$), which follows from Eq. (\ref{eq:VLQ}) with the optimum value of $a = 0$. 
In Fig~\ref{fig:xx_D_sm} (a), (blue, green, red) regions respectively represent allowed parameter space that satisfies the relic abundance $\Omega h^2 = 0.12 \pm 0.005$ for the choice of ratio of mass of $M_{VLL}/M_{VLQ} = (1, 0.3, 0.25)$. Dark (light) brown shaded regions are the exclusion region from $pp \to jj (bb) + {E\!\!\!\!/}_{T}$ obtained by recasting the supersymmetric squark (sbottom) searches. Note that choosing other ratios such as $M_{VLL} > M_{VLQ}$ ($M_{VLL} \lesssim 0.2 M_{VLQ}$) is excluded by squark searches and theoretical limits. Thus, a very narrow parameter space is allowed which could be probed/excluded by future colliders such as HL-LHC and FCC-hh.  

Similarly in Fig~\ref{fig:xx_D_sm} (b), the green band corresponds to the allowed parameter space that satisfies the relic abundance constraint. We fix the VLQ mass at $2.1$ TeV, which satisfies the LHC limits. Heavier VLQ masses would only make the theoretical constraint stronger, reducing the allowed parameter space. Here all the vector-like lepton masses are taken to be degenerate. The red band is the exclusion region from $pp \to l^+ l^- +{E\!\!\!\!/}_{T}$ signature obtained by recasting supersymmetric slepton searches. We find that DM mass from $400-550$ GeV is allowed in this scenario. The thin band where DM mass is close to vector-like fermion mass in both the figures corresponds to the contribution coming dominantly from co-annihilation.         

The parameter space discussed above can be slightly improved by including all the annihilation modes of Fig.~\ref{fig:anni}. We fix the mass of the $T$-even SM singlet Higgs field $h'$ at $(0.3, 0.5)$ TeV such that when $M_{DM} > M_{h'}$, pairs of DM can annihilate into pairs of $h'$ (we keep $\tilde{h}$ field to be heavy here).  Fig.~\ref{fig:combined_plot} shows the allowed parameter space that satisfies relic density including various theoretical and experimental bounds discussed earlier. The dark (light) blue band corresponds to the case when $M_{VLL} = M_{VLQ}$ for the choice of $M_{h'} = 0.5\ (1.0)$ TeV. As is clear from the figure, the allowed range of the DM mass is increased to $\sim 0.9$ TeV compared to $\sim 0.8$ TeV from Fig.~\ref{fig:xx_D_sm}. Similarly, by taking the ratio of $M_{VLL}/M_{VLQ} = 0.3$, an improvement on DM mass from 550 GeV to 670 GeV is achieved as shown by the green band in Fig.~\ref{fig:combined_plot}. We have verified that each of these numerical solutions is consistent with an approximate analytical solution for the relic density given in Ref.~\cite{Berlin:2014tja} for real vector DM at low relative velocity.  We conclude that vector boson DM is consistent within the trinification model when its mass is less than 900 GeV. This corresponds to a theoretical bound on the vector-like quarks of about 4.5 TeV. The unitarity limit on vector-like lepton Yukawa coupling is slightly weaker, which would allow its mass to be as high as about 6 TeV.

Spin-independent direct DM-nucleon cross-section for \cite{Hisano:2010yh, Hisano:2015bma, Barman:2017yzr} can in principle give a stringent constraint on the parameter space of the vector boson DM model. The $t$-channel diagrams which can potentially give large contributions to direct detection cross section are subdominant in the limit of tiny mixing between the SM Higgs and singlet Higgs $h'$. Tree-level diagrams involving vector-like quark exchange can generate interactions of vector boson dark matter with quarks which can constrain the parameter space of the model. However, we find that the cross sections are far below the exclusion limit for almost all the parameter space, except in the $s$-channel when $M_{DM} \sim M_{VLQ}$. This exclusion is depicted as a black band overlayed on top of relic density plot in Fig.~\ref{fig:xx_D_sm} and Fig.~\ref{fig:combined_plot}. Note that the loop contributions to direct detection amplitudes via penguin diagram with the exchange of $Z/\gamma$ and vector boson dark matter coupling to gluons via box diagrams can be important, which we have not 
included in the present work.       


\section{Conclusion}\label{sec:con}

We have developed in this paper plausible dark matter candidates arising from TeV scale trinification theories.  Symmetry breaking can be achieved while preserving a trinification parity ($T$-parity).  Under this parity, the SM fermions are all even, while the vector-like quarks and leptons present in the theory are odd.  Among the new gauge bosons some are $T$-even while some are $T$-odd.  This setup admits doublet-singlet fermionic DM candidate, as well as a scalar singlet DM candidate.

The main focus of the paper has been the identification of a $T$-odd vector boson as the DM candidate.  It has off-diagonal couplings connecting SM fermions and vector-like fermions.  We have analyzed the DM phenomenology of this setup.  LHC searches for dilptons with missing energy as well as dijet with missing energy provide strong constraints on the model parameters.  We have shown consistent parameter space which satisfies these limits while being also consistent with relic abundance and direct DM detection limits. The model predicts the vector boson DM mass to be below 900 GeV, along with upper limits of 4.5 TeV on the vector-like quark masses.  The entire parameter space of the model should be explored in future collider experiments. 
\section*{Acknowledgement}
The work of KSB is supported by the US Department of Energy Grant No. DE-SC 0016013.

\appendix

\section{Appendix}\label{sec:append}

Here we work out explicitly the scalar spectrum of the model in the electroweak symmetry preserving limit as discussed in Sec.~\ref{sec:higgs}. 
There are 36 complex components in the $\chi_{ij}^{\alpha \beta}$ field which decompose into 6 triplets, 6 doublets, and 6 singlets fields under the weak $SU(2)_L$, whereas $\Phi_i^\alpha$ decomposes into 3 doublets and 3 singlet fields. These fields are identified and their mass matrices worked out here. 

\subsection{\boldmath{$T$} parity even fields}

\subsubsection{Weak-triplet $T$-even scalars}
The $T$ parity even weak triplets defined in Eq.~\eqref{eq:Tparity} have the field composition given by
\begin{equation}
    \chi_{EE}\ (1,3,1) =
    \begin{pmatrix}
     \chi_{22}^{11} \\
     \chi_{12}^{11} \\
     \chi_{11}^{11}
    \end{pmatrix} \, , \hspace{5mm} 
    \chi_{EE^c}\ (1,3,1) =
    \begin{pmatrix}
     \chi_{11}^{22*} \\
     \chi_{12}^{22*} \\
     \chi_{22}^{22*}
    \end{pmatrix} \, , \hspace{5mm} 
    \chi_{e\nu}\ (1,3,1) =
    \begin{pmatrix}
     \chi_{11}^{33*} \\
     \chi_{12}^{33*} \\
     \chi_{22}^{33*}
    \end{pmatrix} \, . \hspace{5mm} 
\end{equation}
These states mix to form a $3\times 3$ matrix. 
The elements of the mass-squared matrix $M_{ij}$  read as
\begin{align}
   M_{11} &= -\lambda_{10} V_n^{2}-2(\lambda_3+\lambda_4) V_N^{2}-2(\lambda_2 + \lambda_3+\lambda_4+\lambda_5) V_\nu^2 \nonumber\\
    M_{22}  &= -\lambda_{10} V_n^2 - 2 (\lambda_3 + \lambda_4) V_N^2 - 2 (\lambda_4 + \lambda_5) V_\nu^2  \nonumber \\
   M_{22} &= (\lambda_9 - \lambda_{10}) V_n^2 + 2 (\lambda_2 - \lambda_4) V_N^2 - 2 (\lambda_2 + \lambda_3 + \lambda_4 + \lambda_5) V_\nu^2 \nonumber\\
   M_{12} &= 2 \lambda_{14} V_n^2 + 6 \mu_\chi V_N \nonumber \\
   M_{13} &= 6 \mu_\chi V_\nu \nonumber \\
   M_{23} &= 2 \lambda_{3} V_N V_\nu 
\end{align}
For the benchmark values given in Table~\ref{tab:bound}, these squared masses are all positive and the masses in units of $V_N$ are respectively $\{3.62, 2.80, 0.96\}$ and $\{2.86, 2.42, 1.70\}$ for {\tt BP-I} and {\tt BP-II}. 

\subsubsection{Weak-doublet $T$-even scalars}
The four $T$ parity even doublets have the composition given as follows:
\begin{align}
     \chi_{EN}\ (1,2, 1/2) &= 
    \begin{pmatrix}
     \chi_{23}^{13} \\
     \chi_{13}^{13}
    \end{pmatrix} \, , \hspace{5mm} 
    \chi_{E^cN}\ (1,2,1/2) = 
    \begin{pmatrix}
     \chi_{13}^{23*} \\
     \chi_{23}^{23*}
    \end{pmatrix} \, , \hspace{5mm}  \label{eq:doub4}\\
     \Phi_{EE^c}\ (1,2,1/2) &= 
    \begin{pmatrix}
     \Phi_{1}^{2*} \\
     \Phi_{2}^{2*}
    \end{pmatrix} \, , \hspace{5mm} 
    \Phi_{EE}\ (1,2, 1/2) = 
    \begin{pmatrix}
     \Phi_{2}^{1} \\
     \Phi_{1}^{1}
    \end{pmatrix} \, . \hspace{5mm} 
    \label{eq:doub5}
\end{align}
These states mix to form a $4\times 4$ matrix. The elements of the squared mass matrix, in the basis  $(\chi_{EN},\chi_{E^cN},\Phi_{EE^c}, \Phi_{EE})$,  denoted as $\tilde{M}_{ij}$, read as
\begin{align}
    \tilde{M}_{11} &= \frac{1}{4} (-2 \lambda_{10} + \lambda_{11} + 2 \lambda_9) V_n^2 + (\lambda_2 - 2 \lambda_3 - \lambda_4 + \lambda_5) V_N^2 \nonumber \\
    &~~~- (2 \lambda_2 + 2 \lambda_3 + \lambda_4 + 2 \lambda_5) V_\nu^2 \\
    \tilde{M}_{22} &= \frac{1}{4} (-2 \lambda_{10} + \lambda_{11} + 2 \lambda_9 ) V_n^2 + (\lambda_2 - 2 \lambda_3 - \lambda_4 + \lambda_5) V_N^2 \nonumber\\
    &~~~- (\lambda_2 + 2 \lambda_3 + \lambda_4 + \lambda_5) V_\nu^2 \\
    \tilde{M}_{33} &=  -2 \lambda_7 V_n^2 + (-\lambda_{10} + \lambda_{11} + \lambda_9) V_N^2 + (-\lambda_{10} + \lambda_9) V_\nu^2 \nonumber\\
    &~~~+ \frac{4}{V_n^2} (\lambda_2 + \lambda_5) V_N^2 (V_N^2 - V_\nu^2) \\
    \tilde{M}_{44} &= -2 \lambda_7 V_n^2 - \lambda_{10} (V_N^2 + V_\nu^2) + V_n^2 [ \lambda_{11} + \lambda_9 \\
    &~~~+ \frac{4}{V_n^2} (\lambda_2 + \lambda_5) (V_N^2 - V_\nu^2)] \nonumber\\
    \tilde{M}_{12} &= -\frac{1}{2} \lambda_{12} V_n V_N \\
    \tilde{M}_{13} &= -\lambda_{13} V_n^2 - \lambda_{12} V_N^2 \\
    \tilde{M}_{14} &= -\frac{1}{2} \lambda_{11} V_n V_N - \lambda_9 V_n V_N - \frac{2}{V_n} (\lambda_2 + \lambda_5) V_N (V_N^2-V_\nu^2) \\
    \tilde{M}_{23} &= -\frac{1}{2} \lambda_{11} V_n V_N - \lambda_9 V_n V_N - \frac{2}{V_n} (\lambda_2 + \lambda_5) V_N (V_N^2-V_\nu^2)\\
    \tilde{M}_{24} &= -\lambda_{13} V_n^2 + \lambda_{12} (-V_N^2 + V_\nu^2) \\
    \tilde{M}_{34} &= - V_n (\mu_\phi + 2 \lambda_{13} V_N)~.
\end{align}
The mass-squared values of these $T$-even doublets  are all positive and the masses normalized to $V_N$ for benchmark {\tt BP-I} and {\tt BP-II} are $\{2.86, 2.53, 2.21, 2.02 \}$ and $\{2.72, 2.36, 1.83, 1.64 \}$, respectively. 

\subsubsection{Weak-singlet $T$-even scalars}
There are five $T$-even $SU(2)_L$ singlet fields  identified as
\begin{align}
 \chi_{33}^{11}\ (1,1,2) \, , \hspace{5mm} \chi_{33}^{12}\ (1,1,1) \, , \hspace{5mm} \chi_{33}^{22}\ (1,1,0) \, , \hspace{5mm} \chi_{33}^{33}\ (1,1,0) \, , \hspace{5mm} \Phi_3^3\ (1,1,0) \, .
\end{align}
The first two in this list are doubly charged and singly charged scalars, while the rest are neutral. 
The singly charged field $\chi_{33}^{12}$ is a Goldstone mode. The mass of the doubly charged $\chi_{33}^{11}$ field  is given by
\begin{align}
    M_{\chi_{33}^{11}}^2 = -2 V_\nu^2 (\lambda_2 + \lambda_5)~.
    \label{eq:dobulecharged}
\end{align}
Notice that $(\lambda_2 + \lambda_5) < 0$ is required for this mass parameter to be positive, which is satisfied in the two benchmark points which correspond to mass eigenvalues  of $0.133\ V_N\ (0.118\ V_N)$ for {\tt BP-I} ({\tt BP-II}).

There is mixing among the fields $\{ \text{Re}[\chi_{33}^{22}], \text{Re}[\chi_{33}^{33}], \text{Re}[\Phi_{3}^{3}] \}$  with the mass matrix elements given by 
\begin{align}
    M_{11} &= 4 (\lambda_1 + \lambda_2 + \lambda_3 + \lambda_4 + \lambda_5) V_\nu^2\\
    \label{eq:lightH1}
    M_{22} &= (\lambda_{11} + \lambda_9) V_n^2 + 2 [2 \lambda_1 + 3 \lambda_2 + 2(\lambda_3 + \lambda_4) + 3 \lambda_5] V_N^2 \\
    M_{33} &= 4 (\lambda_6 + \lambda_7) V_n^2 \\
    \label{eq:masshp}
    M_{12} &= 4 (\lambda_1 + \lambda_3 + \lambda_4) V_N V_\nu \\
    M_{13} &= 2 (\lambda_{10} + \lambda_{8} ) V_n V_\nu \\
    M_{23} &= 2 (\lambda_{10} + \lambda_8) V_n V_N - \frac{4 V_N}{V_n} (\lambda_2 + \lambda_5) (V_N^2- V_\nu^2)  ~.
    \label{eq:lightH2}
\end{align}
The eigenvalues of the the above $3\times 3$ mass matrix are positive and the masses normalized to $V_N$ are given as $\{1.30, 0.65, 0.057\}$ and $\{1.60, 0.39, 0.095 \}$ for {\tt BP-I} and {\tt BP-II}, respectively. The lightest field is identified as $h'$, relevant for DM analysis. 
In the limit $Re[\Phi_3^3]$ mixing with $\text{Re}[\chi_{33}^{22}], \text{Re}[\chi_{33}^{33}]$ small, the mass of the $h' = \text{Re}[\Phi_3^3]$ is simply given as
\begin{equation}
    M_{h'} \simeq 2 \sqrt{\lambda_{h'}}\ V_n \, ,
\end{equation}
where $\lambda_{h'} = \lambda_6 + \lambda_7$. This allows the quartic coupling of the interaction $
\frac{\lambda_{h'}}{4}\ (h')^4 $, essential for DM pair annihilation into pairs of $h'$, to be written as (Cf. Eq. (\ref{eq:DMmass}))
\begin{equation}
    \lambda_{h'} = \frac{g_R^2 }{8}\ \frac{M_{h'}^2\ (1+2a^2)}{M_{W_7}^2}~.
    \label{eq:quartichp}
\end{equation}

The neutral field Im$[\chi_{33}^{22}]$ is a Goldstone modes. There is mixing among the neutral field $\{ \text{Im}[\chi_{33}^{33}],  \text{Im}[\Phi_{3}^{3} \}$ yielding one Goldstone and one physics state:
\begin{align}
    G''^0 &= \frac{2 V_N \text{Im}[\chi_{33}^{33}] + V_n \text{Im}[\Phi_{3}^{3}] }{\sqrt{V_n^2 + 4 V_N^2}} \\
    X''^0 &= \frac{- V_n \text{Im}[\chi_{33}^{33}] + 2 V_N \text{Im}[\Phi_{3}^{3}] }{\sqrt{V_n^2 + 4 V_N^2}}
\end{align}
with the mass of the physical field given by
\begin{equation}
    M_{X''^0}^2 = \frac{1}{V_n^2} (V_n^2 + 4 V_N^2) \big[(\lambda_{11} + \lambda_{19}) V_n^2 + 2 (\lambda_2 + \lambda_5) (V_N^2 - V_\nu^2)\big] \, .
    \label{eq:Msig5}
\end{equation}
The mass-squared corresponding to Eq.~\eqref{eq:Msig5} is positive and its mass is  $3.75\ V_N$ for {\tt BP-I}  and $3.80\ V_N$ for {\tt BP-II}. 

\subsection{\boldmath{$T$} parity odd fields}

\subsubsection{Weak-triplet $T$-odd scalars}

There are  three $T$ parity odd $SU(2)_L$ triplet fields identified as follows:
\begin{equation}
    \chi_{Ee}\ (1,3,1) =
    \begin{pmatrix}
     \chi_{11}^{23*} \\
     \chi_{12}^{23*} \\
     \chi_{22}^{23*}
    \end{pmatrix} \, , \hspace{5mm} 
    \chi'_{EE}\ (1,3,0) =
    \begin{pmatrix}
     \chi_{11}^{12*} \\
     \chi_{12}^{12*} \\
     \chi_{22}^{12*}
    \end{pmatrix} \, , \hspace{5mm} 
    \chi'_{E\nu}\ (1,3,0) =
    \begin{pmatrix}
     \chi_{11}^{13*} \\
     \chi_{12}^{13*} \\
     \chi_{22}^{13*}
    \end{pmatrix} \, .
\end{equation}
Here the first triplet carries a hypercharge, and does not mix with the other two fields which have $Y=0$. 
The mass-squared for the $Y=1$ triplet $\chi_{Ee}$  is given by
\begin{align}
     M_{\chi_{Ee}}^2 &= \frac{1}{2} (\lambda_9 - 2 \lambda_{10}) V_n^2 + (\lambda_2 - 2 (\lambda_3 + \lambda_4)) V_N^2 - (\lambda_2 + 2 (\lambda_3 + \lambda_4 + \lambda_5)) V_\nu^2 ~.
\end{align}
The $Y=0$ triplets are complex fields. Its neutral components $\chi_{12}^{12}$ and $\chi_{12}^{13}$ have the following masses:
\begin{align}
    M_{(\text{Re}, \text{Im})[\chi_{12}^{12}]}^2 &= (- \lambda_{10} + 2 \lambda_{14}) V_n^2 \pm 6 \mu_\chi V_N - 2 (\lambda_3 + \lambda_4) V_N^2 - (\lambda_2 + 2 (\lambda_3 + \lambda_4 + \lambda_5) ) V_\nu^2 \, , \label{eq:trip1}\\
     M_{(\text{Re},\text{Im})[\chi_{12}^{13}]}^2 &= \frac{1}{2} (\lambda_9 - 2 \lambda_{10}) V_n^2  \pm 6 \mu_\chi V_\nu + (\lambda_2 - 2 (\lambda_3 + \lambda_4)) V_N^2  - 2 (\lambda_2 + \lambda_3 + \lambda_4 + \lambda_5) V_\nu^2 \, . \label{eq:trip2}
\end{align}
The fields $\{ \chi_{11}^{12*}, \chi_{22}^{12} \}$ and $\{ \chi_{11}^{13*}, \chi_{22}^{13} \}$  mix to form $2\times2$ matrices. The eigenstates corresponding to these matrices are 
\begin{eqnarray}
    X_{\pm} &= \frac{1}{\sqrt{2}} ( \chi_{22}^{12} \pm \chi_{11}^{12*}) \, , \\
     X'_{\pm} &= \frac{1}{\sqrt{2}} ( \chi_{22}^{13} \pm \chi_{11}^{13*}) \, .
\end{eqnarray}
The masses corresponding to the physical state $X_{\mp}$ and $X'_{\mp}$ are respectively given by Eqs.~\eqref{eq:trip1} and ~\eqref{eq:trip2}. 
For the benchmark values given in Table~\ref{tab:bound}, the mass-squared values are positive and the masses for the fields $\{M_{\chi_{Ee}},\ M_{\text{Re}[\chi_{12}^{12}]},\ M_{\text{Im}[\chi_{12}^{12}]},\ M_{\text{Re}[\chi_{12}^{13}]},\ M_{\text{Im}[\chi_{12}^{13}]} \}$ noramalized to $V_N$ are respectively $\{2.80, 1.81, 3.50, 1.98, 3.43 \}$ and $\{2.42, 1.96, 2.77, 2.03, 2.75\}$ for {\tt BP-I} and {\tt BP-II}. 
\subsubsection{Weak-doublet $T$-odd scalars}
There is one $(1,2,3/2)$ multiplet that is $T$ odd, as well as four $(1,2,1/2)$  multiplets. The $Y=3/2$ doublet is identified as
\begin{align}
     \chi_{Ee^c}\ (1,2,3/2) &= 
    \begin{pmatrix}
     \chi_{23}^{11} \\
     \chi_{13}^{11}
    \end{pmatrix} \, 
\end{align}
with its mass given by
\begin{equation}
     M_{\chi_{Ee^c}}^2 = -\frac{1}{2} \lambda_{10} V_n^2 - (2\lambda_3 + \lambda_4) V_N^2 -(2 \lambda_2 + 2 \lambda_3 + \lambda_4 + 2 \lambda_5) V_\nu^2~.
\end{equation}
This squared mass is positive and the mass eigenvalue is $2.19 \,V_N$ for {\tt BP-I} and $1.80\, V_N$ for {\tt BP-II}.
The four $T$ parity odd doublets with $Y=1/2$ that mix with each other are
\begin{align}
     \chi_{E\nu^c}\ (1,2, 1/2) &= 
    \begin{pmatrix}
     \chi_{23}^{12} \\
     \chi_{13}^{12}
    \end{pmatrix} \, , \hspace{5mm} 
    \chi_{E^c\nu^c}\ (1,2,1/2) = 
    \begin{pmatrix}
     \chi_{13}^{22*} \\
     \chi_{23}^{22*}
    \end{pmatrix} \, , \hspace{5mm}  \label{eq:doub2}\\
     \chi_{eN}\ (1,2,1/2) &= 
    \begin{pmatrix}
     \chi_{13}^{33*} \\
     \chi_{23}^{33*}
    \end{pmatrix} \, , \hspace{5mm} 
    ~\Phi_{e\nu}\ (1,2, 1/2) = 
    \begin{pmatrix}
     \Phi_{1}^{3*} \\
     \Phi_{2}^{3*}
    \end{pmatrix} \, . \hspace{5mm} 
    \label{eq:doub3}
\end{align}
The elements of this mass matrix, in the basis  $(\chi_{E\nu^c}^{i},\chi_{E^c\nu^c}^i,\chi_{eN}^i, \Phi_{e\nu}^i)$, denoted as $\hat{M}_{ij}$, are given by
\begin{align}
    \hat{M}_{11} &= -\frac{1}{2} \lambda_{10} V_n^2 -(2 \lambda_3 + \lambda_4) V_N^2 - (\lambda_2 + 2 \lambda_3 + \lambda_4 + \lambda_5) V_\nu^2 \\
    \hat{M}_{22} &= -\frac{\lambda_{10}}{2} V_n^2 - 2 (\lambda_3 + \lambda_4) V_N^2\\
    \hat{M}_{33} &= \frac{V_n^2}{2} (-\lambda_{10} + \lambda_{11} + 2 \lambda_9)  + 2 (\lambda_2 + \lambda_5) V_N^2 - (2 \lambda_2 + 2 \lambda_3 + \lambda_4 + 2 \lambda_5) V_\nu^2 \\
    \hat{M}_{44} &= - \lambda_{10} (V_N^2 + V_\nu^2) + V_N^2 [\lambda_{11} + 2 \lambda_9 + \frac{4}{V_n^2} (\lambda+2 + \lambda_5) (V_N^2 - V_\nu^2) ]\\
    \hat{M}_{12} &= -\frac{\lambda_{12}}{\sqrt{2}} V_n V_\nu\\
    \hat{M}_{13} &= 0 \\
    \hat{M}_{14} &= \lambda_{12} V_\nu^2\\
    \hat{M}_{23} &= (2 \lambda_3 + \lambda_4) V_N V_\nu\\
    \hat{M}_{24} &= \frac{\lambda_{10}}{\sqrt{2}} V_n V_\nu \\
    \hat{M}_{34} &= \frac{V_n V_N}{\sqrt{2}} [ \lambda_{10} - \lambda_{11} - 2 \lambda_9 - \frac{4}{V_n^2} (\lambda_2 + \lambda_5) (V_N^2 - V_\nu^2)]~.
\end{align}

There are two Goldstone modes associated with the doublets fields given in Eq.~\eqref{eq:doub2} and Eq.~\eqref{eq:doub3}:
\begin{align}
    G^+ &= \frac{\sqrt{2} V_\nu \chi_{13}^{22*} + \sqrt{2} V_N \chi_{13}^{33*} + V_n \Phi_{1}^{3*}}{\sqrt{V_n^2 + 2 (V_N^2 + V_\nu^2)}}  \, , \\
    G^0 &=\frac{\sqrt{2} V_\nu \chi_{23}^{22} + \sqrt{2} V_N \chi_{23}^{33} + V_n \Phi_{2}^{3}}{\sqrt{V_n^2 + 2 (V_N^2 + V_\nu^2)}}  \, .
\end{align}
The masses of the remaining $T$ parity odd $(1,2,1/2)$ fields for are $\{2.93, 2.21, 2.10 \} V_N$  for {\tt BP-I} and $\{2.78, 1.83, 1.71 \} V_N$ for {\tt BP-II}, with all eigenvalues being positive. 


\subsubsection{Weak-singlet $T$-odd scalars}
There are four complex $T$-odd singlet fields, which are identified as
\begin{align}
     \chi_{33}^{13}\ (1,1,1) \, ,  \hspace{5mm} \chi_{33}^{23}\ (1,1,0) \, , \hspace{5mm} \Phi_3^1\ (1,1,1)\, ,  \hspace{5mm}  ~~\Phi_3^2\ (1,1,0)  \, .
\end{align}
The singly charged singlet fields $\{ \chi_{33}^{13}, \Phi_{3}^1 \}$ mix to give one physical state and a Goldstone mode:
\begin{align}
    G'^+ &= \frac{\sqrt{2} V_N \chi_{33}^{13} + V_n \Phi_3^1}{\sqrt{V_n^2 + 2 V_N^2}} \\
    X'^+ &= \frac{- V_n \chi_{33}^{13} + \sqrt{2} V_N \Phi_3^1}{\sqrt{V_n^2 + 2 V_N^2}}
\end{align}
The mass of the physical state $X'^+$ is  given by
\begin{equation}
    M_{X'^+}^2 = \frac{1}{2 V_n^2} (V_n^2 + 2 V_N^2) [(\lambda_{11} + \lambda_9) V_n^2 + 4 (\lambda_2 + \lambda_5) (V_N^2 -V_\nu^2)]~.
\end{equation}
This mass is $ 1.84\ V_N\ (1.88\ V_N)$ for {\tt BP-I} ({\tt BP-II}).

The neutral singlet fields $\{\text{Re}[ \chi_{33}^{23}], \text{Re}[\Phi_{3}^2] \}$ mix to give one Goldstone mode and one physical state:
\begin{align}
    G'^{0r} &= \frac{\sqrt{2} (V_N-V_\nu) \text{Re}[\chi_{33}^{23}] + V_n \text{Re}[\Phi_3^2]}{\sqrt{V_n^2 + 2 |V_N- V_\nu|^2}} \\
    X'^{0r} &= \frac{- |V_N-V_\nu| V_n \text{Re}[\chi_{33}^{23}] + \sqrt{2} |V_N-V_\nu| \text{Re}[\Phi_3^2]}{\sqrt{V_n^2 + 2 |V_N- V_\nu|^2}}~.
\end{align}
The mass of the physical state $X'^{0r}$ is given by
\begin{equation}
    M_{X'^{0r}}^2 = \frac{1}{2 V_n^2} \big[V_n^2 + 2 (V_N - V_\nu)^2\big] \big[(\lambda_{11} + \lambda_9) V_n^2 + 4 V_N (\lambda_2 + \lambda_5)  (V_N + V_\nu)\big] \, .
    \label{eq:Msig3}
\end{equation}

As noted earlier, from the doubly charged scalar mass, $\lambda_2 + \lambda_5 < 0$ is required, which forces the choice $\lambda_2 + \lambda_5 = b\ (\lambda_9+ \lambda_{11})\ V_n^2/V_N^2$ for the positivity of the mass term in  Eq.~\eqref{eq:Msig3}, with $b$ being  order one or smaller. The mass of the neutral $T$-odd field ($\equiv H' \sim (1,1,0)$) is $ 0.165\ V_N$ for {\tt BP-I} and $0.187\ V_N$ for {\tt BP-II} with positive eigenvalue.
Similarly, the neutral fields $\{\text{Im}[ \chi_{33}^{23}], \text{Im}[\Phi_{3}^2] \}$ mix to give one Goldstone mode and one physical state:
\begin{align}
    G'^{0i} &= \frac{\sqrt{2} (V_N+V_\nu) \text{Im}[\chi_{33}^{23}] + V_n \text{Im}[\Phi_3^2]}{\sqrt{V_n^2 + 2 |V_N + V_\nu|^2}} \\
    X'^{0i} &= \frac{- |V_N+V_\nu| V_n \text{Im}[\chi_{33}^{23}] + \sqrt{2} |V_N+V_\nu| \text{Im}[\Phi_3^2]}{\sqrt{V_n^2 + 2 |V_N + V_\nu|^2}}~.
\end{align}
The mass of the physical state  is given by
\begin{equation}
    M_{X'^{0i}}^2 = \frac{1}{2 V_n^2} \big[V_n^2 + 2 (V_N + V_\nu)^2\big] \big[(\lambda_{11} + \lambda_9) V_n^2 + 4 (\lambda_2 + \lambda_5) V_N (V_N - V_\nu)\big] \, .
    \label{eq:Msig4}
\end{equation}
The squared mass corresponding to Eq.~\eqref{eq:Msig4} is  positive with the mass being $3.51\ V_N$ for {\tt BP-I} and $3.60\ V_N$ for {\tt BP-II}.

\bibliographystyle{utphys}
\bibliography{references}

\end{document}